\pdfoutput=1
\documentclass[11pt,a4paper]{article}

\usepackage[T1]{fontenc}
\usepackage[utf8]{inputenc}
\usepackage[a4paper,hmargin=1.1in,vmargin=1.15in]{geometry}
\usepackage{amsmath}
\usepackage{newtxtext,newtxmath}
\usepackage{graphicx}
\usepackage{booktabs}
\usepackage[font=small,labelfont=bf,labelsep=period]{caption}
\usepackage[authoryear]{natbib}
\usepackage{xcolor}
\usepackage[colorlinks=true,allcolors=linkblue]{hyperref}

\definecolor{linkblue}{rgb}{0.05,0.2,0.55}
\hypersetup{
  pdftitle={The Akhmediev breather for the Dysthe equation and the unstable manifold of the Stokes wave},
  pdfauthor={David Andrade}}

\bibpunct{(}{)}{;}{a}{}{,}
\numberwithin{equation}{section}

\graphicspath{{figures/}}
\newcommand{\ii}{\mathrm{i}}
\newcommand{\ee}[1]{\mathrm{e}^{#1}}
\newcommand{\dd}{\mathrm{d}}
\newcommand{\Wu}{W^{\mathrm{u}}}
\newcommand{\Eu}{E^{\mathrm{u}}}
\newcommand{\eps}{\varepsilon}
\newcommand{\Nw}{N_{\mathrm{w}}}
\newcommand{\kc}{k_0}
\newcommand{\wc}{\omega_0}
\newcommand{\Kmod}{K}
\newcommand{\kone}{k_1}
\newcommand{\thone}{\theta_1}
\newcommand{\lamone}{\lambda_1}
\newcommand{\Peak}{P}
\newcommand{\abs}[1]{\lvert#1\rvert}
\newcommand{\Tr}{\mathcal{T}}

\newcommand{\Real}{\mbox{Re}}

\makeatletter
\renewcommand{\@seccntformat}[1]{\csname prefix@#1\endcsname\csname the#1\endcsname\quad}
\newcommand{\backmatterhead}[1]{\par\addvspace{\medskipamount}\noindent\textbf{#1}\enspace\ignorespaces}
\makeatother

\title{The Akhmediev breather for the Dysthe equation\\ and the unstable manifold of the Stokes wave}

\author{David Andrade\\[6pt]
  {\normalsize School of Sciences and Engineering, Universidad del Rosario, 111711 Bogot\'a, Colombia}\\[2pt]
  {\normalsize\href{mailto:davideu.andrade@urosario.edu.co}{\texttt{davideu.andrade@urosario.edu.co}}}}

\date{}

\begin{document}

\maketitle

\begin{abstract}
In this article we define and compute the breather of the Dysthe equation for deep water waves, the analogue of the Akhmediev breather of the nonlinear Schr\"odinger equation (NLS). When a single pair of modes is unstable, the unstable manifold of the Stokes wave consists of the Stokes wave and the translates and time shifts of a single solution, the breather.  The construction uses only gauge and translation invariance, not integrability or a Hamiltonian structure.  We use the parameterisation method to find the breather as a power series whose coefficients follow recursively from linear equations. This is not an asymptotic expansion: the series is the Taylor series of the breather itself.  For deep water waves we map the envelope amplification over steepness and group length and give a simple formula, based on the linear stability analysis, that summarises it to within a few per cent.  Our results agree well with Euler simulations: the edge of the instability band of the Dysthe equation lies much closer to the onset of growth than that of the NLS, and, away from the onset and below breaking, the crest and height amplifications lie less than $5\%$ below the Euler ones.
\end{abstract}

\section{Introduction}\label{sec:intro}

A uniform train of waves on deep water, a Stokes wave, is unstable. Small modulations of its amplitude grow at the expense of the carrier wave, and the train breaks up into groups. This is the Benjamin--Feir, or modulational, instability \citep{benjamin_feir1967,zakharov1968}. In the nonlinear Schr\"odinger equation (NLS) the whole process, from a small modulation to the largest wave and back to the uniform train, is described exactly by the Akhmediev breather \citep{akhmediev_korneev1986,akhmediev_etal1987}. Out of an orderly wave train, a breather produces a wave much larger than the waves around it, and for this reason breathers are regarded as prototypes of freak waves. They have also been observed in wave tanks \citep{chabchoub_etal2011,chabchoub_etal2012}.

The NLS is, however, only the leading-order model, and at the steepness of real wave groups breathers depart from it. In the experiments of \citet{shemer_alperovich2013} the spectrum of the Peregrine breather widened asymmetrically, which the NLS cannot reproduce, and the measurements agreed instead with the Dysthe equation. \citet{slunyaev_shrira2013} followed breathers in simulations of the Euler equations and found that the NLS describes them only qualitatively: some modulations that it predicts to grow do not grow at all, and the crests grow higher than its envelope. The next model in the hierarchy is the equation of \citet{dysthe1979}, which keeps the terms of the next order in the steepness, among them the mean flow induced by the wave group, and \citet{trulsen_etal2000} improved it further with the exact linear dispersion. These models are not integrable, and they do not share the reflection symmetry $X\mapsto-X$ of the NLS.

Here is where the difficulty begins. Without integrability there is no formula for the breather, and it is not even clear what the breather of the Dysthe equation is. In practice it is computed by perturbing the Stokes wave along its unstable mode, with a small amplitude, and integrating the equations forward in time, as \citet{slunyaev_shrira2013} did for the Euler equations. Such initial data lie close to the breather, but not on it. Already for the NLS they lead to periodic orbits that approach the breather only as the perturbation tends to zero, and data on the two sides of the breather evolve differently \citep{ablowitz_herbst1990,calini_etal1996}. The outcome therefore depends on how the computation is started, and the error grows with the size of the initial perturbation (figure~\ref{fig:FS}).

In this paper we take a different route, and the paper has two main parts. The first part is mathematical. We define the breather through the unstable manifold of the Stokes wave, that is, the set of solutions that converge to the Stokes wave exponentially fast as time tends to $-\infty$. It turns out that two symmetries are enough to describe this manifold completely: the gauge invariance, which turns the Stokes wave into a fixed point in a rotating frame, and the invariance under translations along the wave train. Whenever a single pair of eigenvalues is unstable, the unstable manifold is a surface made of the Stokes wave and of the translates and time shifts of a single solution, which we call the breather. The breather is therefore unique, and its peak is a property of the model and of its parameters, not of an initial condition.

We show this for the Galerkin truncations of the models, and find numerically that their breathers converge as the truncation is increased. Nothing else is needed: integrability, a Hamiltonian structure, the reflection symmetry and time reversibility play no part. For the NLS the breather is the Akhmediev breather, in agreement with the integrable theory \citep{ablowitz_herbst1990,li_mclaughlin1994}.

To compute the breather we adapt the parameterisation method \citep{cabre_etal2003,haro_etal2016,reinhardt_mj2019} to our setting. The Stokes wave is not a hyperbolic fixed point, but it has a single positive, unstable eigenvalue $\lambda_1$, which is what the method needs, and the translations reduce the computation to a single real variable, $\xi=\ee{\lambda_1\tau}$. The breather becomes a power series in $\xi$ whose coefficients are vectors of Fourier amplitudes. They are found recursively, to any order, from linear systems that can always be solved.

We stress that this is not an asymptotic expansion, which would merely approximate the breather. For the NLS the series converges, it is the Taylor series of the breather itself, and inside its disc of convergence it is the breather. We find this disc exactly: its boundary passes through the peak of the Akhmediev breather, so the series describes the whole growth of the wave up to its peak. For the Dysthe equations the series converges for every truncation of the model; whether it converges for the full equations is an open question, but we expect it to. To go beyond the disc we integrate the equations from a point inside it, which lies on the unstable manifold and not merely close to it. Against the Akhmediev breather, the computed peak is correct to $2\times10^{-8}$. The leading coefficients of the series also give, in every model, the instability cascade that \citet{chin_etal2015} described for the NLS.

The second part of the paper is about deep water waves. We take the Dysthe equation with the exact dispersion as our model and compute the peak $\Peak$ of its breather, the largest value of the envelope, over the whole instability band and for steepness up to $0.12$. Since this breather need not return to the Stokes wave, we define its first peak through the exchange of wave action between the carrier and the sidebands. The resulting map is our main physical result. It shows that $\Peak$ lies below the value of the NLS over most of the band, and rises above it only for the longest groups of the steepest wave trains. The map is summarised, to within $3\%$ for steepness up to $0.1$, by the formula of the Akhmediev breather, $1+2\sin\thone$, evaluated with the angle $\thone$ of the unstable eigenvector of the Dysthe equation. The formula needs only the linear stability of the Stokes wave, so it is easy to use, but the fact that it holds is a result of the nonlinear computations. From the envelope we also reconstruct the free surface. The highest crest is that of a local Stokes wave of amplitude $\Peak$, with no fitted constants, while the wave height amplification stays below $\Peak$.

Finally, we compare our breathers with the Euler simulations of \citet{slunyaev_shrira2013}. The edge of the instability band of the Dysthe equation lies much closer to the onset of growth in the Euler equations than that of the NLS. For groups of six or more waves, once the crests have grown well above the background and until they approach breaking, our crest and height amplifications lie less than $5\%$ below theirs. \citet{slunyaev_shrira2013} described these features by fitting the NLS; here they follow from the Dysthe equation with no fitted constants.

This paper is organised as follows. In section~\ref{sec:models} we present the envelope models, the reconstruction of the free surface and the models in Fourier space. In section~\ref{sec:stokes} we study the linear stability of the Stokes wave, describe its unstable manifold and the breather, and define its first peak. In section~\ref{sec:construction} we construct the breather as a power series, find its radius of convergence for the NLS and continue it beyond. In section~\ref{sec:deepwater} we discuss the consequences for trains of deep water waves: the breathers of the four models, the map of $\Peak$ and its linear summary, the crest and height amplifications, and the comparison with the Euler equations. Finally, section~\ref{sec:conclusions} summarises our results. Appendix~\ref{sec:numerics} gives the numerical details, appendix~\ref{sec:firstorders} the first orders of the recursion, and appendix~\ref{sec:ABfourier} the Akhmediev breather and its Fourier modes in our variables.

\section{Trains of deep water waves}\label{sec:models}

We consider unidirectional, long-crested trains of deep-water waves represented as a carrier wave modulated by a slowly varying envelope $B$ which depends on a single horizontal coordinate $x$.  The carrier has a wavenumber $\kc$ and an amplitude $a_0$, so it has frequency $\wc=\sqrt{g\kc}$ and steepness $\eps=\kc a_0$.  To leading order the free surface elevation is
\begin{equation}\label{eq:surface}
  \eta(x,t)=\Real\bigl[B\,\ee{\ii (k_0x - \omega_0t)}\bigr],
\end{equation}
and we work throughout with
\begin{equation}\label{eq:vars}
  u=\frac{\bar B}{a_0},\qquad
  X=\sqrt2\,\eps \kc\,(x-c_g t),\qquad
  \tau=\tfrac12\eps^2\wc t,\qquad c_g=\frac{\wc}{2\kc},
\end{equation}
so that a uniform wave train has $\abs{u}=1$ and $\tau$ runs forward in physical time.

The conjugation in \eqref{eq:vars} gives the forward time orientation, but it also makes the $n$th Fourier mode of $u$,  denoted by $nK$,  correspond to the surface wavenumber $\kc-n\Kmod$, so positive $n$ is the \emph{lower} sideband of the carrier $\kc$.

All envelopes under consideration are periodic in $X$ and so each of them has a discrete Fourier spectrum defined by a \emph{modulation wavenumber}, which is denoted by $K$ in physical units and $k_1$ in the dimensionless ones. They are related through
\begin{equation}\label{eq:K}
  \frac{\Kmod}{\kc}=\sqrt2\,\eps\kone.
\end{equation}
The corresponding period of the envelope is therefore $2\pi/k_1$.

It is often clearer to use, instead of $K$,  the number of carrier waves per modulation, $\Nw=\kc/\Kmod$.  The wave dynamics depend on two parameters only: the wave steepness $\eps$, and either $\kone$ or $\Nw$.

\subsection{Envelope models for deep water}\label{sec:threemodels}

All four models used here are of the form
\begin{equation}\label{eq:class}
  \ii\,\partial_\tau u=\Omega(-\ii\partial_X)u-F(u),\qquad
  F(u)=u\bigl(1-s\abs{\partial_X}\bigr)\abs{u}^2+\gamma\abs{u}^4u
       -\ii c\bigl(p\abs{u}^2u_X+q\,u^2\bar u_X\bigr),
\end{equation}
with $\Omega$ a real Fourier multiplier and $\abs{\partial_X}$ the multiplier $\abs{k}$. The coefficients are
collected in table~\ref{tab:models}.
\begin{table}
  \centering
  \begin{tabular}{@{}lccccccl@{}}
    \toprule
    model & $\Omega(k)$ & $s$ & $\gamma$ & $c$ & $p$ & $q$ & order \\
    \midrule
    NLS & $k^2/2$ & $0$ & $0$ & $0$ & --- & --- & $O(\eps^3)$ \\
    Stokes-corrected & $k^2/2$ & $0$ & $\eps^2$ & $0$ & --- & --- & $O(\eps^3)+O(\eps^5)$ \\
    Dysthe & $k^2/2+(\sqrt2\eps/4)k^3$ & $\sqrt2\,\eps$ & $0$ & $-\sqrt2\,\eps$ & $3$ & $1/2$ & $O(\eps^4)$ \\
    Dysthe, exact dispersion & eq.~\eqref{eq:exactdisp} & $\sqrt2\,\eps$ & $0$ & $-\sqrt2\,\eps$ & $3$ & $1/2$ & $O(\eps^4)$ \\
    \bottomrule
  \end{tabular}
  \caption{The models of section~\ref{sec:threemodels} as coefficient sets of \eqref{eq:class}, in the variables
    \eqref{eq:vars} where the background is $\abs{u}=1$ and $\tau$ runs forward. The Dysthe equation with
    the exact dispersion keeps only the modes $n<\Nw$, whose surface wavenumber $\kc-n\Kmod$ is positive. All are written for the
    surface-elevation envelope; the velocity-potential and Zakharov conventions differ in $q$, taking
    $-1/2$ and $0$, and each carries its own reconstruction of the free surface. The last column is the
    order in steepness to which the model is derived, not the order of the terms it retains: the
    Stokes-corrected equation keeps one term of $O(\eps^5)$ while omitting those of $O(\eps^4)$.}
  \label{tab:models}
\end{table}

Within the class of models in \eqref{eq:class} we have:

\textbf{The NLS:} This model corresponds to setting $s=c=\gamma=0$ and $\Omega=k^2/2$. It is the leading-order, narrow-band, envelope equation for deep water wave trains \citep[see][]{zakharov1968,stiassnie1984}. The model is of order $O(\eps^3)$, has a Hamiltonian structure and is integrable.

\textbf{The Stokes-corrected NLS:} This model adds the quintic term $\gamma=\eps^2$, which reproduces the fifth-order frequency correction $\omega=\wc(1+\tfrac12\eps^2+\tfrac12\eps^4)$ of a Stokes wave \citep{fenton1985}. It is not a consistent deep-water model, since it keeps a term of $O(\eps^5)$ while discarding those of $O(\eps^4)$. However, it is the simplest non-integrable model of the class that keeps the reflection symmetry of the NLS; whenever $u(X,\tau)$ is a solution, so is $u(-X,\tau)$.  In \eqref{eq:class} this symmetry holds when $c=0$ and $\Omega$ is even.

\textbf{The Dysthe equation:} This model adds every term of $O(\eps^4)$. These include the induced mean flow, which appears in \eqref{eq:class} as the nonlocal multiplier $s\abs{\partial_X}$ acting on $\abs{u}^2$ with $s=\sqrt2\,\eps$,  the third-order dispersion, $\Omega(k)=k^2/2+(\sqrt2\eps/4)k^3$,  and the two cubic derivative terms, with $p=3$, $q=\tfrac12$ and a common coefficient $c=-\sqrt2\,\eps$. These are the coefficients of the surface elevation envelope \citep[eq.~19]{trulsen_etal2000}.  \citet{dysthe1979} wrote the equation for the envelope of the velocity potential instead,  and the change of variable reverses the sign of $q$.   In this model the terms carrying $c$ and the cubic dispersion are odd under $X\mapsto-X$, so the Dysthe equation breaks the reflection symmetry of the NLS models.

\textbf{The Dysthe equation with exact dispersion:} This model replaces the cubic polynomial by the exact multiplier
\begin{equation}\label{eq:exactdisp}
  \Omega(k)=\frac{2}{\eps^2}\Bigl[1-\frac{\kappa}{2}-\sqrt{1-\kappa}\Bigr],\qquad
  \kappa=\sqrt2\,\eps k,
\end{equation}
which is the deep-water dispersion relation of the sideband  $\kc-n\Kmod$ written in the variables and the moving frame of \eqref{eq:vars}. The two dispersions agree to order $O(\eps^2)$.

We prefer the exact dispersion relation to its Taylor approximation because, in deep water,  shorter waves are always slower. Consider, for example,  a group of shorter surface waves near the wavenumber $\kc(1-\kappa)$ ($\kappa<0$ corresponds to surface wavenumbers larger than $k_0$).  With the exact dispersion its group velocity is $c_g/c_{g0}=(1-\kappa)^{-1/2}$ so,  in the frame of the carrier group, groups of shorter waves drift backwards.  With the Taylor dispersion the group velocity is $c_g/c_{g0}=1+\kappa/2+3\kappa^2/8$, which returns to one at $k=7\kc/3$,  so every group of waves shorter than that travels faster than the carrier group and drifts forwards.

The exact multiplier \eqref{eq:exactdisp} removes this at no cost but it is real only for $\kappa<1$, so we pose the model on the modes $n<\Nw$ alone. The free surface then contains only right-going waves as $\kc-n\Kmod>0$. The modes with $n\ge\Nw$ would be waves of zero or negative wavenumber.

The truncation also follows the Zakharov equation, from which the envelope models descend. In one horizontal dimension the interaction kernel vanishes for non-trivial quartets that have one negative and three positive (or three negative and one positive) wavenumbers (see \citealp[p.~703]{dyachenko_zakharov2011}), so at this order right-propagating waves do not generate left-propagating ones.  The kernel also vanishes when one of the wavenumbers is zero \citep[appendix~A]{andrade_stuhlmeier2023}.  The nonlinear terms in the envelope models come from expansions about the carrier of the interaction kernel of the Zakharov equation and they do not inherit these properties.  Posing the model on $n<\Nw$, where every surface wavenumber is positive, restores them.

\subsection{Reconstruction of the free surface}\label{sec:reconstruction}

The models evolve the envelope, not the free surface itself.  This can be reconstructed from the \emph{free waves} and their \emph{bound harmonics}.  At Dysthe's order the reconstruction of \citet[appendix~C]{gramstad_trulsen2011}, written for the elevation envelope with $\chi=\wc t-\kc x$ and $s=\sqrt2\eps$, is
\begin{equation}\label{eq:recon}
  \frac{\eta}{a_0}=\Real\bigl[u\ee{\ii\chi}\bigr] -\frac{\sqrt2}{4}\eps^2\abs{\partial_X}\abs{u}^2
  +\frac{\eps}{2}\Real\bigl[(u^2+\ii s\,uu_X)\ee{2\ii\chi}\bigr]
  +\frac38\eps^2\Real\bigl[u^3\ee{3\ii\chi}\bigr].
\end{equation}
The first term is the free wave. It is followed by the mean-level set-down that accompanies the group, the second and third harmonics.  For the NLS the consistent reconstruction keeps only the free wave and the second harmonic, with $s=0$. For the Stokes-corrected equation, the third harmonic is also kept.

\subsection{Models in Fourier space}

Fix a modulation wavenumber $k_1$. The envelope $u$ is periodic in $X$ with period $2\pi/k_1$, so it can be written as a Fourier series
\begin{equation}\label{eq:fourier}
  u(X,\tau)=\sum_{n = -\infty}^{\infty}A_n(\tau)\,\ee{\ii nk_1X}.
\end{equation}
Here $A_n$ is the complex amplitude of the envelope wavenumber $nk_1$, which corresponds to the surface wavenumber $\kc-n\Kmod$.

Substituting \eqref{eq:fourier} into \eqref{eq:class} turns each model of table~\ref{tab:models} into an infinite system of ordinary differential equations,
\begin{equation}\label{eq:fouriermodels}
  \ii\frac{\dd A_n}{\dd\tau}=\Omega(nk_1)A_n
  -\sum_{j,l,m}\delta_{n,j}^{l,m}\,T_{n,j}^{l,m}\,\bar A_jA_lA_m
  -\gamma\!\!\sum_{j,j',l,m,m'}\!\!\delta_{n,j,j'}^{l,m,m'}\,\bar A_j\bar A_{j'}A_lA_mA_{m'},
\end{equation}
where $\delta_{n,j}^{l,m}$ equals one when $n+j=l+m$ and zero otherwise.  Likewise $\delta_{n,j,j'}^{l,m,m'}$ equals one when $n+j+j'=l+m+m'$, and vanishes in any other case. The kernel is given by
\begin{equation}\label{eq:kernel}
  T_{n,j}^{l,m}=1-\frac{sk_1}{2}\bigl(\abs{n-l}+\abs{n-m}\bigr)+ck_1\Bigl(\frac p2\,(l+m)-q\,j\Bigr).
\end{equation}
Its three parts are the cubic nonlinearity, which contributes the $1$; the induced mean flow, which gives the terms proportional to $s$ and acts through the separation of the interacting wavenumbers, because $\abs{\partial_X}$ acts on $\abs{u}^2$, whose Fourier modes are the differences $l-j$; and the two derivative terms, which give the terms in $c$, $p$ and $q$ and act through the wavenumbers of the Fourier modes they differentiate.

For the NLS $T_{n,j}^{l,m}=1$ and $\gamma=0$. The Stokes-corrected equation keeps $T_{n,j}^{l,m}=1$ and adds the quintic sum with $\gamma=\eps^2$. For the Dysthe equation $s=-c=\sqrt2\,\eps$, and since $\sqrt2\,\eps k_1=\Kmod/\kc=1/\Nw$ by \eqref{eq:K}, the coefficient takes the form
\begin{equation}\label{eq:dysthekernel}
  T_{n,j}^{l,m}=1-\frac{\abs{n-l}+\abs{n-m}+3n+2j}{2\Nw},
\end{equation}
in which $l+m=n+j$ has been used.

Dispersion enters only through $\Omega(nk_1)$. For the exact multiplier \eqref{eq:exactdisp}, $\Omega(nk_1)=(2/\eps^2)\bigl[1-\tfrac{n}{2\Nw}-(1-\tfrac{n}{\Nw})^{1/2}\bigr]$ is real only for $n<\Nw$, and the sums in \eqref{eq:fouriermodels} run over $n,j,l,m<\Nw$ and $n,j,j',l,m,m'<\Nw$.

\subsubsection{Truncated models}

For numerical simulations the models must be truncated. We do so by restricting every index in \eqref{eq:fouriermodels} to $\abs{n}\le N$. For the Dysthe equation with the exact dispersion the truncation acts on its upper sidebands only, and the modes kept are $-N\le n\le\min(N,\lceil\Nw\rceil-1)$; for $N<\Nw$ this is again $\abs{n}\le N$.

Two properties of the full and truncated systems can be read off term by term, and they carry the whole construction of the breather. Every term has one more $A$ than $\bar A$, so the system is equivariant under the gauge action, that is, $A_n\mapsto\ee{\ii\varphi}A_n$; and every term obeys $n+j=l+m$, so it is equivariant under $A_n\mapsto\ee{\ii nk_1a}A_n$, which is the translation $X\mapsto X+a$. The same holds for the quintic sum.

All the models conserve the wave action, which is  given by
\begin{align}
M = \sum_n\abs{A_n}^2,
\end{align}
and for their truncation it is the finite sum $M_N=\sum_n\abs{A_n}^2$ with $|n|\leq N$. The kernel \eqref{eq:kernel} has the symmetry $T_{n,j}^{l,m}=T_{l,m}^{n,j}$ of a Hamiltonian system exactly when $cq=0$, and only then is the corresponding model Hamiltonian in these variables, with the momentum $\sum_nnk_1\abs{A_n}^2$ also conserved. The Dysthe equation is not Hamiltonian and does not conserve  momentum \citep[p.~404]{gramstad_trulsen2011}.

\section{Linear stability of a Stokes wave}\label{sec:stokes}

A \emph{Stokes wave} is a solution whose only nonzero Fourier mode is $n=0$. All the models above, full or truncated, have them, and a simple computation shows that the Stokes waves are
\begin{align}
A_0(\tau) = b\,\ee{\ii\left(|b|^2 + \gamma|b|^4\right)\tau},\qquad\text{for some $b\in\mathbb{C}$.}
\end{align}
The phase of $b$ shifts the Stokes wave in time, so we set it to zero. The modulus of $b$ is determined by the total wave action, which we normalise to 1. So,  without any loss of generality we take the Stokes wave to be $A_0(\tau) = \ee{\ii\left(1 + \gamma\right)\tau}$.

We now switch to a frame that rotates with the frequency of the Stokes wave and, in anticipation of the linear stability analysis, moves with a velocity $V$ to be chosen later. That is, we write
\begin{align}
u=\ee{\ii(1+\gamma)\tau}\sum_nA_n\ee{\ii nk_1(X-V\tau)},
\end{align}
where the Fourier modes now satisfy
\begin{equation}\label{eq:galerkin}
\begin{split}
  \ii\frac{\dd A_n}{\dd\tau}={}&\bigl[\Omega(nk_1)+1+\gamma-Vnk_1\bigr]A_n\\
  &-\sum_{j,l,m}\delta_{n,j}^{l,m}\,T_{n,j}^{l,m}\,\bar A_jA_lA_m
  -\gamma\!\!\sum_{j,j',l,m,m'}\!\!\delta_{n,j,j'}^{l,m,m'}\,\bar A_j\bar A_{j'}A_lA_mA_{m'}.
\end{split}
\end{equation}
Note that in this frame of reference the Stokes wave becomes a stationary solution with $A_0 = 1$ and $A_n = 0$ for $n\neq 0$.

\subsection{Linear stability analysis}

We consider a small perturbation about the Stokes wave of the form $A_0=1+a_0$ and $A_n=a_n$.  Then, we substitute it in equation \eqref{eq:galerkin} and keep the terms linear in $a$. In the cubic sum they come from the index triples in which two of $j$, $l$, $m$ vanish, namely $(j,l,m)=(0,n,0)$, $(0,0,n)$ and $(-n,0,0)$,  and in the quintic sum from their analogues, so that
\begin{equation}\label{eq:linear}
  \ii\frac{\dd a_n}{\dd\tau}=\alpha_na_n-\beta_n\bar a_{-n},\qquad
  \alpha_n=\Omega(nk_1)-Vnk_1+1-2T_{n,0}^{n,0}-2\gamma,\qquad
  \beta_n=T_{n,-n}^{0,0}+2\gamma .
\end{equation}
The coefficient $T_{n,-n}^{0,0}$ of $\beta_n$ comes from products of the form $\bar A_{-n}A_0A_0$, so it couples the unperturbed carrier amplitude, to leading order, to the sidebands.

Equation \eqref{eq:linear} couples $a_n$ to $\bar a_{-n}$ only. Each pair obeys
\begin{equation}\label{eq:pair}
  \ii\frac{\dd}{\dd\tau}\begin{pmatrix}a_n\\ \bar a_{-n}\end{pmatrix}
  =\begin{pmatrix}\alpha_n & -\beta_n\\ \beta_{-n} & -\alpha_{-n}\end{pmatrix}
   \begin{pmatrix}a_n\\ \bar a_{-n}\end{pmatrix}.
\end{equation}

The eigenvalues of this system are
\begin{equation}\label{eq:eig}
  \lambda_n^{\pm}=-\frac{\ii}{2}\,(\alpha_n-\alpha_{-n})
  \pm\Bigl[\beta_n\beta_{-n}-\tfrac14(\alpha_n+\alpha_{-n})^2\Bigr]^{1/2}.
\end{equation}

For the two NLS models we have
\begin{align}
 \alpha_n=\frac{n^2k_1^2}{2}-Vnk_1 - 1-2\gamma,\qquad
  \beta_n=1 + 2\gamma.
\end{align}

For the Dysthe models, from \eqref{eq:dysthekernel}, they are
\begin{equation}\label{eq:dysthelin}
  \alpha_n=\Omega(nk_1)-Vnk_1-1+\frac{\abs{n}+3n}{\Nw},\qquad
  \beta_n=1-\frac{2\abs{n}+n}{2\Nw}.
\end{equation}

\subsubsection{Mode $n = 0$}

For the mode $n = 0$, $\alpha_0 = -\beta_0$ and equation \eqref{eq:linear} yields a double zero eigenvalue. The general solution is
\begin{align}
a_0(\tau) = c_1 + \ii(c_2 - 2\alpha_0c_1\tau),
\end{align}
for $c_1$ and $c_2$ some real constants.  Hence a perturbation at the mode $n=0$ is expected to either grow linearly in time or to remain constant.

\subsubsection{Modes $|n| = 1$}

At $|n| = 1$ we choose $V$ so that $\alpha_1 = \alpha_{-1}$, which removes the imaginary part of $\lambda_1^\pm$ in \eqref{eq:eig}. For the NLS, with or without the Stokes correction, $\alpha_{1} - \alpha_{-1} = -2k_1V$, so $V = 0$.

For the Dysthe equations we choose
\begin{equation}\label{eq:drift}
  V=\frac{\Omega_{\mathrm{o}}(k_1)}{k_1}-cp,
\end{equation}
where $\Omega_{\mathrm{o}}(k)=\tfrac12[\Omega(k)-\Omega(-k)]$ is the odd part of the dispersion relation.

With this choice the eigenvalues are
\begin{align}
\lambda_1^{\pm} = \pm\Bigl[\beta_1\beta_{-1}-\tfrac14(\alpha_1+\alpha_{-1})^2\Bigr]^{1/2}.
\end{align}

We write $\lambda_1 = \lambda_1^{+}$ and call it the growth rate.  For the NLS this is the Benjamin--Feir growth rate $\lambda_1=\tfrac12 k_1(4-k_1^2)^{1/2}$, provided that $k_1<2$, see \citet{mei_etal2005}. It is the same for every truncation that contains the modes $0$ and $\pm 1$.

The eigenspace of $\lambda_1$ has complex dimension 1 and, in the variables $(a_1,\bar a_{-1})$ of \eqref{eq:pair}, it is spanned by
\begin{equation}
  w^+=\bigl(\alpha_1+\ii\lambda_1,\;\beta_{-1}\bigr).
\end{equation}
In polar form, we write
\begin{equation}\label{eq:theta}
  \alpha_1+\ii\lambda_1=(\beta_1\beta_{-1})^{1/2}\,\ee{2\ii\thone},\qquad 0<\thone<\tfrac\pi2.
\end{equation}
As we shall see, this angle sets the envelope amplification.

Multiplying $w^+$ by the nonzero scalar $2\beta_{-1}^{-1/2}\ee{-\ii\thone}$, and writing the resulting eigenvector in terms of the Fourier modes $a_1$ and $a_{-1}$ yields
\begin{equation}\label{eq:eigvec}
  a_1=2\beta_1^{1/2}\,\ee{\ii\thone},\qquad a_{-1}=2\beta_{-1}^{1/2}\,\ee{\ii\thone}.
\end{equation}
For the NLS \eqref{eq:theta} gives $\cos\thone=\tfrac12k_1$ and $\lambda_1=\sin2\thone$, and \eqref{eq:eigvec} is $a_{\pm1}=2\ee{\ii\thone}=k_1+\ii(4-k_1^2)^{1/2}$: the angle $\thone$ of the Akhmediev breather is the phase of the unstable eigenvector.

This eigenvector is given by \citet{ablowitz_herbst1990}, whose NLS is ours with time $t=\tau/2$: their unstable direction $(k_1^2+2\ii\lambda_1)\cos k_1X$ is $\tfrac12k_1$ times the perturbation $a_1\ee{\ii k_1X}+a_{-1}\ee{-\ii k_1X}$ of \eqref{eq:eigvec}, and they show that the breather leaves the plane wave along it. They, and \citet{calini_schober2002}, parametrise the breather by the angle $\phi$ with $k_1=2\sin\phi$, so that $\phi=\tfrac\pi2-\thone$.

\subsubsection{Modes $|n|\geq 2$ and the instability band}

For $|n|\geq 2$ the eigenvalues \eqref{eq:eig} depend on the model and on $k_1$. We define the \emph{instability band} as the range of $k_1$ in which $\lambda_1^\pm$ are real and nonzero, while every $\lambda_n^{\pm}$ with $\abs{n}\ge2$ is nonzero and purely imaginary.

For the NLS the instability band is $1<k_1<2$. Adding the Stokes correction changes the instability band to $\sqrt{1+2\eps^2}<k_1<2\sqrt{1+2\eps^2}$. For the Dysthe equation with polynomial dispersion the band is $\tfrac12\sqrt{4+6\eps^2}-\sqrt2\,\eps<k_1<\sqrt{4+6\eps^2}-2\sqrt2\,\eps$ and for the Dysthe equation with the exact dispersion the band is found numerically from \eqref{eq:eig}.

Truncation does not change the instability band, provided $N\ge2$, but some care is needed when handling the Dysthe equation with exact dispersion.  Whenever $n\le-\Nw$, the mode $a_n$ loses its partner $\bar a_{-n}$. In this case its linearisation reduces even further to a single equation of the form
\begin{align}
\ii\frac{\dd}{\dd\tau}a_n=\alpha_na_n,
\end{align}
which always has a single imaginary eigenvalue $-\ii\alpha_n$.

\subsection{The unstable manifold and the breather}\label{sec:manifold}

In this section we look at the definition of a breather, the central concept of this article,  establish its main defining characteristics and its intimate connection with the linear instability of the Stokes wave.  

First of all, note that all the models in \eqref{eq:galerkin} are equivariant under translations $\Tr_a:A_n\mapsto\ee{\ii nk_1a}A_n$, which simply moves the envelope by $a$ along $X$, and gauge transformations $A_n\mapsto\ee{\ii\varphi}A_n$. They all have the following invariant subspace 
\begin{align}
\Sigma_0 =\bigl\{A_n\;:\; \ A_n =0\ \text{for}\ n\ne0,\; A_0\in\mathbb{C}\,\bigr\},
\end{align}
which is also invariant under translations and gauge transformations, and contains all possible Stokes waves.

Moving to the reference frame that rotates with frequency $1 + \gamma$, where \eqref{eq:galerkin} is defined, makes every point on the unit circle of $\Sigma_0$ a stationary solution of \eqref{eq:galerkin}. We concentrate on the point $A_0 = 1$, $A_n = 0$ for $n\neq 0$, and call it simply the Stokes wave, but the following discussion applies to every other point on this circle.

The spectrum at the Stokes wave always has a zero eigenvalue of algebraic multiplicity 2,  which comes from the mode $n=0$: the constant solution $a_0=\ii c_2$ is tangent to the circle of Stokes waves, while $a_0=c_1-2\ii\alpha_0c_1\tau$ points to the Stokes waves of other amplitudes, which rotate relative to the frame. With $\kone$ in the instability band we have only two real eigenvalues $\pm\lambda_1$, the rest being purely imaginary pairs.  We take  $\lambda_1$ to be the only eigenvalue with positive real part. The Stokes wave is therefore  not a hyperbolic fixed point, and the classical unstable manifold theorem for hyperbolic fixed points does not apply directly.

The eigenspace of $\lambda_1$ is
\begin{align}\label{eq:Eu}
  \Eu=\bigl\{\,a_1=2\beta_1^{1/2}\ee{\ii\thone}\zeta,\ \ a_{-1}=2\beta_{-1}^{1/2}\ee{\ii\thone}\bar\zeta,\ \ a_n=0\ \text{for}\ \abs{n}\ne1\;:\;\zeta\in\mathbb{C}\,\bigr\},
\end{align}
and the translation $\Tr_a$ acts on it as a rotation $\zeta\mapsto\ee{\ii k_1a}\zeta$. In this setting the \emph{unstable manifold} $\Wu$ of the Stokes wave is the set of solutions of \eqref{eq:galerkin} that converge to it exponentially fast as $\tau\to-\infty$.

Since $\lambda_1$ is the eigenvalue of largest real part, the non-resonance conditions of the theorem of \citet[theorem 1.1 and remark 6]{cabre_etal2003}, applied backwards in time to the flow of any truncation of \eqref{eq:galerkin}, hold automatically.  The theorem guarantees the existence of an invariant analytic surface of real dimension 2, tangent to $\Eu$ at the Stokes wave and the only invariant surface of class $C^2$ with that tangent, on which the flow is conjugate to the linear flow $\dd\zeta/\dd\tau=\lambda_1\zeta$ on $\Eu$. That surface is exactly $\Wu$, because a solution that converges exponentially to the Stokes wave as $\tau\to-\infty$ can only do so along $\Eu$.

On $\Wu$ there are two types of solutions: the Stokes wave, which is just a point, and the solutions that leave it along $\Eu$. Each of the latter can be labelled by a single $\zeta\ne0$, and as $\tau\to-\infty$ it is of the form
\begin{align}\label{eq:leave}
A_1=2\beta_1^{1/2}\ee{\ii\thone}\zeta\,\ee{\lambda_1\tau},\qquad
A_{-1}=2\beta_{-1}^{1/2}\ee{\ii\thone}\bar\zeta\,\ee{\lambda_1\tau},\qquad
A_0=1,\qquad
A_n=0\ \text{for}\ \abs{n}\ge2,
\end{align}
up to terms of order $\ee{2\lambda_1\tau}$, that is, the Stokes wave plus $\ee{\lambda_1\tau}$ times the element $\zeta$ of \eqref{eq:Eu}. The solution that corresponds to $\zeta = 1$ is \emph{the breather}.

By uniqueness of $\Wu$, any translation $\Tr_a$ maps $\Wu$ onto itself, hence it maps the solution labelled $\zeta$ onto the one labelled $\ee{\ii k_1a}\zeta$. Multiplying $\zeta$ by $r>0$ simply shifts the solution in time by $\lambda_1^{-1}\ln r$, which yields a solution that still converges exponentially to the Stokes wave as $\tau\to-\infty$. Hence any solution on $\Wu$, aside from the Stokes wave, can be obtained by a translation and a time shift of the breather: the one labelled $\zeta=r\ee{\ii k_1a}$ is the breather translated by $a$ and shifted in time by $\lambda_1^{-1}\ln r$. Thus the two real dimensions of $\Wu$ are the time along the breather, set by the modulus of $\zeta$, and its position along the wave train, set by the argument of $\zeta$. Any quantity that is invariant under translations and time shifts, such as the height $\Peak$ of the first peak \eqref{eq:firstevent}, therefore takes a single value on $\Wu$: it is a property of the model and its parameters, not of an initial condition.

We point out that the existence of the breather and its structural characteristics need only the gauge and translation invariance, the single unstable eigenvalue and the uniqueness of $\Wu$.  Integrability, a Hamiltonian structure, other spatial symmetries (such as the reflection $X\mapsto-X$) and time reversibility play no part.

For the NLS the breather is the Akhmediev breather \citep{akhmediev_korneev1986,akhmediev_etal1987}, and the same description of $\Wu$ follows from the integrable theory. \citet{ablowitz_herbst1990} identified the breather as the homoclinic orbit of the unstable mode of the plane wave, using translation invariance to reduce the unstable directions to one (their eq. 9). \citet[sections 4.2--4.3]{li_mclaughlin1994} represented the unstable manifold of the circle of Stokes waves by a B\"acklund transformation whose complex parameter plays the role of $\zeta$: its modulus sets the origin of time and its argument the position of the solution in $X$ \citep[see also][ch.~3]{li2004}. They also noted that the orbit returns to the circle with its phase shifted, by $4\thone$ in our variables. What integrability adds is that $\Wu$ closes up, i.e. the stable and unstable manifolds of the Stokes waves coincide \citep[section 3.2]{calini_etal1996}. Under perturbations of the equation the two manifolds persist but need not coincide, and Mel'nikov methods decide whether a homoclinic orbit survives \citep{calini_etal1996,li_etal1996}.

Two special cases connect with earlier work. When the model is reflection symmetric, as in the NLS models,  the even solutions $A_{-n}=A_n$ form an invariant subspace, which meets $\Wu$ in a single curve, the solutions with $\zeta$ real in \eqref{eq:leave}: the Stokes wave, the breather ($\zeta>0$) and its translate by half a period ($\zeta<0$). This is the setting of the integrable studies above, in which evenness leaves two branches of the manifold \citep[section 4.3]{li_mclaughlin1994}; the Dysthe equations have no such reduction. 

The second case occurs in the truncation $N=1$ of the NLS,  where \eqref{eq:galerkin} is the three-wave system of \citet{trillo_wabnitz1991}. Its phases enter only through $\arg A_1+\arg A_{-1}-2\arg A_0$, which the gauge transformations and the translations leave unchanged, and its reduced phase plane has the breather as a separatrix, as was also observed for the Zakharov equation by \citet[section 5.1.1]{andrade_stuhlmeier2023}.

To compute $\Wu$ we use the parameterisation method \citep{cabre_etal2003,haro_etal2016}, in the Fourier--Taylor form of \citet{reinhardt_mj2019}. It looks for a map $K$ from $\Eu$ to the space of Fourier amplitudes whose image is $\Wu$ and which turns the linear flow on $\Eu$ into the flow of \eqref{eq:galerkin}, that is, $\tau\mapsto K(\ee{\lambda_1\tau}\zeta)$ solves \eqref{eq:galerkin} for every $\zeta$. Differentiating with respect to $\tau$ turns this into an equation for $K$ alone, the \emph{invariance equation} \citep[eq.~7]{reinhardt_mj2019}, which is completed by asking that $K(\zeta)$ is the Stokes wave plus the element $\zeta$ of \eqref{eq:Eu}, to first order in $\zeta$ \citep[eq.~8]{reinhardt_mj2019}. Translation invariance gives $K(r\ee{\ii k_1a})=\Tr_aK(r)$, so $K$ is needed only on the half-line $\zeta=\xi\ge0$, where $K(\ee{\lambda_1\tau})$ is a parametrisation of the breather itself and the invariance equation is simply \eqref{eq:galerkin} with $\dd/\dd\tau$ replaced by $\lambda_1\xi\,\dd/\dd\xi$.

We solve it with a power series in $\xi$ whose coefficients are vectors of Fourier amplitudes. Matching powers of $\xi$ gives the \emph{cohomological equations} of \citet[section~2.2.2]{haro_etal2016}, and this is where the linearisation enters: at order $p$ the new coefficient solves a linear system whose matrix is $p\lambda_1$ minus the linearisation at the Stokes wave and whose right-hand side involves only the lower orders. Since no $p\lambda_1$ with $p\ge2$ is an eigenvalue, every coefficient is determined uniquely. The series converges in some disc $\abs{\xi}<\varrho$; beyond that the breather is continued by integrating \eqref{eq:galerkin}, which is how \citet[section~2.4.2]{haro_etal2016} globalise the manifolds they compute. Section~\ref{sec:construction} carries out both steps.

\subsection{The first peak of the breather}\label{sec:firstpeak}

For the NLS the breather is a homoclinic orbit: it leaves the Stokes wave and returns to it. Along it, $\max_X\abs{u}$ rises from $1$ at $\tau = -\infty$ to a single maximum and returns to $1$ as $\tau \to \infty$.

Let us denote by $\Pi=\abs{A_0}^2/M$ the fraction of the wave action left in the carrier, the rest being spread over the sidebands. Recall that the total wave action $M$ is conserved in all the models. As the peak forms, at time $\tau_*$, the carrier gives its action to the sidebands and $\Pi$ falls; after the peak the process reverses and the action flows back to the carrier, so $\Pi$ rises again.

We take this exchange of wave action as the working definition of the first peak since it does not presuppose that the orbit is homoclinic to the Stokes wave.

Let $\tau_{\min}$ be the first local minimum of $\Pi$ and $\tau_{\mathrm{ret}}$ be the first local maximum after it. The first peak, or the first focusing event, lies in the window
$\tau\le\tau_{\mathrm{ret}}$, and its height is
\begin{equation}\label{eq:firstevent}
  \Peak=\max_{\tau\le\tau_{\mathrm{ret}}}\;\max_X\abs{u(X,\tau)},
\end{equation}
attained at $\tau_*$.

\section{Construction of the breather}\label{sec:construction}

Applying the parameterisation method of section~\ref{sec:manifold}, we introduce the new variable $\xi = \ee{\lambda_1\tau}$ and look for the breather as a power series
\begin{equation}\label{eq:series}
  A_n=\sum_{p\ge0}a_{n,p}\,\xi^p.
\end{equation}
The coefficients $a_{n,p}$ are found recursively, for all $n$ and all $p$, as solutions of linear equations, which we derive next.

The Stokes wave occurs at $\xi=0$, which corresponds to $\tau\to-\infty$, so $a_{0,0}=1$ and $a_{n,0}=0$ for $n\ne0$.

Since $\dd A_n/\dd\tau=\lambda_1\xi\,\dd A_n/\dd\xi$, equation \eqref{eq:galerkin} becomes
\begin{equation}\label{eq:galerkinxi}
\begin{split}
  \ii\lambda_1\xi\frac{\dd A_n}{\dd\xi}={}&\bigl[\Omega(nk_1)+1+\gamma-Vnk_1\bigr]A_n\\
  &-\sum_{j,l,m}\delta_{n,j}^{l,m}\,T_{n,j}^{l,m}\,\bar A_jA_lA_m
  -\gamma\!\!\sum_{j,j',l,m,m'}\!\!\delta_{n,j,j'}^{l,m,m'}\,\bar A_j\bar A_{j'}A_lA_mA_{m'},
\end{split}
\end{equation}
and, $\xi$ being real, $\bar A_n=\sum_p\bar a_{n,p}\xi^p$.  Substituting \eqref{eq:series} and collecting the coefficient of $\xi^p$, the sums contribute products of coefficients whose orders add up to $p$.  Of those products, those that contain a coefficient of order $p$ must have every other factor of order zero. Since $a_{m,0}=0$ for $m\ne0$, they vanish unless every other factor is $a_{0,0}=1$. Together with the first term of \eqref{eq:galerkinxi}, they reproduce the linear terms in \eqref{eq:linear}. We denote the rest by $R_{n,p}$; it involves only the orders $1$ to $p-1$. Order $p$ of \eqref{eq:galerkinxi} is then
\begin{equation}\label{eq:homological}
  (\ii p\lambda_1-\alpha_n)\,a_{n,p}+\beta_n\,\bar a_{-n,p}=-R_{n,p},
\end{equation}
and together with the complex conjugate of the same equation for $-n$ it is a $2\times2$ system for
$(a_{n,p},\bar a_{-n,p})$,
\begin{equation}\label{eq:block}
  \begin{pmatrix}\ii p\lambda_1-\alpha_n & \beta_n\\ \beta_{-n} & -\ii p\lambda_1-\alpha_{-n}\end{pmatrix}
  \begin{pmatrix}a_{n,p}\\ \bar a_{-n,p}\end{pmatrix}
  =-\begin{pmatrix}R_{n,p}\\ \bar R_{-n,p}\end{pmatrix},
\end{equation}
whose determinant is $(p\lambda_1-\lambda_n^{+})(p\lambda_1-\lambda_n^{-})$, by \eqref{eq:eig}.   A mode without its partner, e.g. when $n\le-\Nw$ in the Dysthe equation with the exact dispersion, has \eqref{eq:homological} without its second term, a single equation whose determinant $\ii p\lambda_1-\alpha_n$ has modulus at least $p\lambda_1$.

The whole construction rests on that determinant. At order $p=1$ the right-hand side vanishes and a nonzero solution exists only where the determinant does, at $n = \pm1$.  For $p\ge2$ it never vanishes. Every $\lambda_n^\pm$ has real part at most $\lambda_1$, being imaginary for $\abs{n}\ne1$ and equal to $\pm\lambda_1$ for $n=\pm1$, so every factor of the determinant satisfies $\abs{p\lambda_1-\lambda_n^\pm}\ge(p-1)\lambda_1\ge\lambda_1$. Each coefficient beyond the first order is therefore determined uniquely by the ones below it, and since this bound depends neither on $n$ nor on $p$, the recursion can be continued to every order and every $n$.

In the language of section~\ref{sec:manifold}, \eqref{eq:galerkinxi} is the invariance equation on the half-line $\zeta=\xi\ge0$, \eqref{eq:block} are its cohomological equations, and \eqref{eq:series} is the Taylor series of the parameterisation, which converges in a disc $\abs{\xi}<\varrho$ and is unique once the eigenvector is fixed. Note that this construction applies to any of the models in their full or truncated versions alike.

\subsection{Recursion}
To illustrate the recursion, we give the first order here; the next two are worked out in appendix~\ref{sec:firstorders}.

\subsubsection{First order}

With $R_{n,1}=0$, \eqref{eq:block} has the nonzero solution
\begin{equation}\label{eq:a1}
  a_{\pm1,1}=2\beta_{\pm1}^{1/2}\,\ee{\ii\thone},\qquad a_{n,1}=0\ \text{ for }\abs{n}\ne1,
\end{equation}
which is the unstable eigenvector \eqref{eq:eigvec}, as expected.

Any multiple $r\,a_{\pm1,1}$ with $r>0$ also solves \eqref{eq:block}. It multiplies every coefficient $a_{n,p}$ by $r^p$, that is, it replaces $\xi$ by $r\xi$, which only shifts the origin of $\tau$ by $\lambda_1^{-1}\ln r$, as in section~\ref{sec:manifold}. We fix $r=1$. For the NLS, \eqref{eq:a1} reduces to $a_{\pm1,1}=2\ee{\ii\thone}$.

\subsubsection{Higher orders}

The second and third orders, in appendix~\ref{sec:firstorders}, show the pattern that holds at every order and for every model of the class: $a_{n,p}=0$ unless $\abs{n}\le p$ and $n\equiv p\pmod2$. Both conditions follow by induction on $p$, because in each product of \eqref{eq:galerkinxi}, cubic or quintic, the harmonics of the factors add, with a minus sign for the conjugated ones, and so do their orders. Harmonic $n$ therefore enters at order $\xi^{\abs{n}}$, driven by products of the harmonics below it.

The following counting gives the dependence on the truncation. Each order changes the harmonic by at most one, so a coefficient $a_{m,i}$, which requires $i\ge\abs{m}$, can influence $a_{n,p}$ only if $\abs{m}-\abs{n}\le p-i$, that is, only if $2\abs{m}\le\abs{n}+p$. The truncation removes the harmonics $\abs{m}\ge N+1$ (for the exact dispersion, the upper ones, $m\le-N-1$), so $a_{n,p}$ takes its untruncated value as soon as $N\ge(\abs{n}+p)/2$, the carrier coefficients are exact up to $p=2N$, and the leading coefficients $a_{\pm n,n}$ need only $N\ge n$.

\subsubsection{The leading coefficients and the instability cascade}

The first coefficient that appears in the power series expansion of the mode $A_{\pm n}$ is  $a_{\pm n,n}$.  Its value depends only on the diagonal terms $a_{\pm i,i}$ with $i<n$. For the NLS, they can be computed to any order in closed form to get
\begin{equation}\label{eq:diagonal}
  a_{\pm n,n}=2\ee{\ii\thone}\sec^{n-1}\thone,\qquad n\ge1,
\end{equation}
a geometric progression with ratio $\sec\thone=2/k_1$.

These diagonal coefficients form the \emph{cascading instability} described by \citet{chin_etal2015}, in which the instability of the modes $n=\pm1$ forces every higher mode to grow in lockstep with them.  What is remarkable about it is that every higher mode is linearly stable, so that on its own a perturbation of it would not grow.  Moreover they show that, as $\tau\to-\infty$, $\abs{A_{\pm n}}=C_n\ee{n\lambda_1\tau}$, so that $\ln\abs{A_n}$ is linear in $\abs{n}$,  which is what they call the triangular spectrum.  From the exact amplitudes they compute the prefactor $C_n$. In our variables it is $C_n = 2\cos\thone\sin^{\abs{n}}\thone$.  

In appendix~\ref{sec:ABfourier} we write the Akhmediev breather of \citet{chin_etal2015} in our variables and compute its Fourier modes, which are of the form
\begin{equation}\label{eq:abmodes}
  A_n=F\,q^{\abs{n}}\quad(n\ne0),\qquad A_0=F-\ee{2\ii\thone},
\end{equation}
with $\varrho=\sin\thone\cos\thone$, which section~\ref{sec:radius} shows to be the radius of convergence of the series, and 
\begin{equation}\label{eq:abmodesxi}
\begin{gathered}
  F=2\cos\thone\,\ee{\ii\thone}\,\frac{(\varrho^2+\ee{2\ii\thone}\xi^2)^{1/2}}{(\varrho^2+\ee{-2\ii\thone}\xi^2)^{1/2}},\\
  q=\frac{2\varrho\sin\thone\,\xi}{\varrho^2+\xi^2+(\varrho^2+\ee{2\ii\thone}\xi^2)^{1/2}(\varrho^2+\ee{-2\ii\thone}\xi^2)^{1/2}}.
\end{gathered}
\end{equation}  

Now we show that the leading coefficients of the power series, i.e. $a_{\pm n,n}$, give this cascade exactly. As $\xi\to0$, $q=\xi\sec\thone+O(\xi^3)$ and $F=2\cos\thone\,\ee{\ii\thone}+O(\xi^2)$ by \eqref{eq:abmodesxi}, so that
\begin{equation}\label{eq:cascade}
  A_{\pm n}=2\cos\thone\,\ee{\ii\thone}\,(\xi\sec\thone)^n+O(\xi^{n+2})
           =a_{\pm n,n}\,\xi^n+O(\xi^{n+2}),\qquad n\ge1,
\end{equation}
with $a_{\pm n,n}$ the diagonal coefficients \eqref{eq:diagonal}. By \eqref{eq:xiab}, $\abs{a_{\pm n,n}}\xi^n=2\cos\thone\,(\sin\thone\,\ee{\lambda_1\tau})^n=C_n\ee{n\lambda_1\tau}$, which is the cascading instability of the Akhmediev breather with its prefactor included. The series also gives the phase: every $a_{\pm n,n}$ carries the factor $\ee{\ii\thone}$, the phase of the unstable eigenvector, which locks the phase of each mode to that of $A_1$, as noted by \citet[eq.~33]{chin_etal2015}.

\begin{figure}
  \centering
  \includegraphics[width=\textwidth]{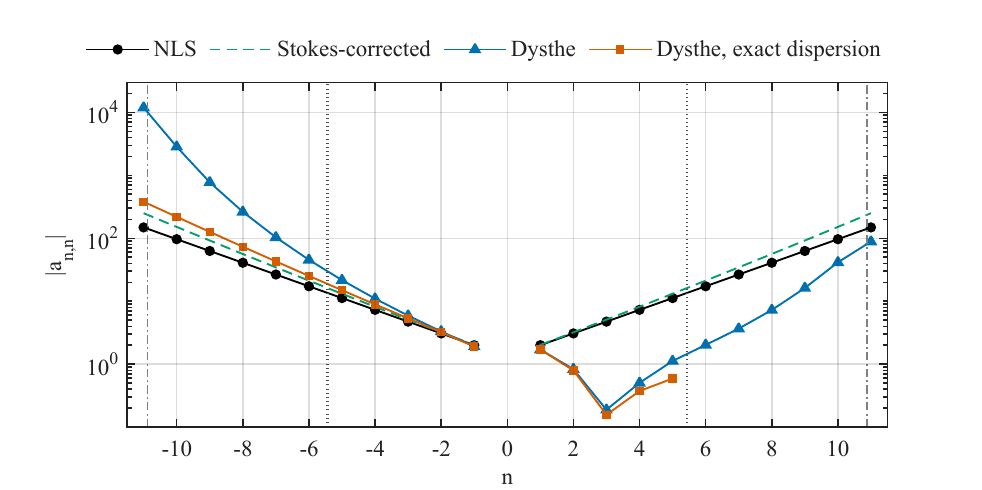}
  \caption{The diagonal terms $\abs{a_{\pm n,n}}$ at $\eps=0.1$ and $k_1=1.3$ ($\Nw=5.4$), for the four models of table~\ref{tab:models}. The NLS coefficients \eqref{eq:diagonal} are the same on both sides. The dotted lines mark $\abs{n}=\Nw$, beyond which the Dysthe equation with the exact dispersion has no modes ($n\ge\Nw$). The dash-dotted lines mark $\abs{n}=2\Nw$; at $n=-2\Nw$ the Taylor dispersion of the Dysthe equation vanishes. Computed with $N=J=12$, so that every coefficient shown is free of truncation error.}
  \label{fig:FL}
\end{figure}

Figure \ref{fig:FL} shows the instability cascade for each model.  The cascade for the NLS with or without the Stokes correction is symmetric and its terms grow geometrically, so they lie on a straight line on a logarithmic axis, the so-called triangular spectrum of the cascading instability. The symmetry of the cascade corresponds to the $X\to-X$ symmetry shared by both models.

The Dysthe equations have asymmetric cascades. The difference between the two sides comes from the terms that break the reflection symmetry.  We shall encounter this asymmetry again in the breather of the Dysthe equation, but it is already present in the leading coefficients of the series, before the series is summed.

Neither Dysthe model is integrable, and their leading coefficients do not form a geometric progression. Moreover, for the Dysthe model with the exact dispersion the right end of the cascade stops before reaching $\Nw$, as the model excludes those modes.

For the Dysthe models, the left terms of the cascade (negative $n$) grow faster than those of the NLS, whereas for positive $n$, they decay and reach a minimum before growing again, and the reason is the kernel. The forcing $R_{\pm n,n}$ of a leading coefficient $a_{\pm n,n}$, $n\ge1$, contains only products of leading coefficients, $\bar a_{\mp i,i}\,a_{\pm l,l}\,a_{\pm m,m}$ with $l+m=n-i$ and $i,l,m\ge0$, and by \eqref{eq:dysthekernel} each of them carries the factor (take $n>0$)
\begin{equation}\label{eq:kerneldiag}
  T_{n,-i}^{l,m}=1-\frac{4n-i}{2\Nw}\quad\text{(mode $n$)},\qquad
  T_{-n,i}^{-l,-m}=1+\frac{2n-3i}{2\Nw}\quad\text{(mode $-n$)},
\end{equation}
whatever the split of $n-i$ into $l$ and $m$. For the modes $-n$ this kernel is larger than one for the terms with $i=0$, and indeed for all $i<2n/3$, which enhances the cascade. For the modes with $n>0$, the first factor in \eqref{eq:kerneldiag} decreases with $n$ and vanishes at $n=\Nw/2+i/4$, so the zeros of the kernels of the different terms, $0\le i\le n-1$, together with that of $\beta_n=1-3n/(2\Nw)$, which couples $a_{n,n}$ to the upper sideband through the block \eqref{eq:block}, are spread over $\Nw/2\le n\le2\Nw/3$. Across that window the forcing $R_{\pm n,n}$ nearly cancels and changes sign, and $\abs{a_{n,n}}$ passes through a minimum, at $n=3$ for $\Nw=5.4$ in figure~\ref{fig:FL}; beyond it the factors grow in magnitude and the cascade grows again.

Our approach therefore gives the instability cascade of the breather in every model, including those that have fewer symmetries than the NLS, are not integrable or are not Hamiltonian. The same holds for the truncated models, where the cascade stops at $N$.

\subsection{Radius of convergence of the Akhmediev breather}\label{sec:radius}

For the NLS the exact breather is known as a function of $\xi$, \eqref{eq:abxi}, and it solves the recursion \eqref{eq:homological} exactly. The series \eqref{eq:series}, whose coefficients are found through \eqref{eq:homological}, is therefore the Taylor series of the Akhmediev breather in powers of $\xi$ about $\xi = 0$.

Its radius of convergence can be read off \eqref{eq:abmodesxi}. For $\abs{\xi}<\varrho$ the arguments $\varrho^2+\ee{\pm2\ii\thone}\xi^2$ have positive real part, so the two square roots are analytic there. The denominator of $q$ does not vanish either, because $(\varrho^2+\xi^2)^2-(\varrho^2+\ee{2\ii\thone}\xi^2)(\varrho^2+\ee{-2\ii\thone}\xi^2)=4\varrho^2\sin^2\thone\,\xi^2$, so it could only vanish at $\xi=0$, where it equals $2\varrho^2$. Hence every mode is analytic in the disc $\abs{\xi}<\varrho$. On its boundary lie the zeros of $\varrho^2+\ee{\pm2\ii\thone}\xi^2$, the four branch points
\[
  \xi=\pm\ii\varrho\,\ee{\pm\ii\thone}=\pm\varrho\,\ee{\pm\ii(\pi/2-\thone)},
\]
and at the two of them where $\varrho^2+\ee{-2\ii\thone}\xi^2=0$ the factor $F$, and with it every mode, becomes infinite. The radius of convergence of the series, which is the same for every Fourier mode, is therefore
\begin{equation}\label{eq:radius}
  \varrho=\sin\thone\cos\thone=\tfrac12\lambda_1.
\end{equation}
Note that a factor $r$ on the eigenvector, as in \eqref{eq:a1}, would divide it by $r$.

By \eqref{eq:xiab} the peak of the breather lies at $\xi=\sin\thone\cos\thone$, exactly on the circle of convergence. At $\xi=\varrho\,\ee{\lambda_1\tau}$, the circle $\abs{\xi}=\varrho$ is the line $\Real\tau=0$ in complex time, and the singularities of the breather lie at the imaginary times $\tau=\pm\ii(\pi/2-\thone)/\lambda_1$ and $\tau=\pm\ii(\pi/2+\thone)/\lambda_1$, modulo $2\pi\ii/\lambda_1$, directly above the instant of the peak. The series therefore converges on the whole growth towards the peak, $\tau<0$, and not beyond it. Reaching the peak, the point of maximum amplification, requires the continuation of section~\ref{sec:continuation}. The limitation is a property of the breather, whose singularities in complex time bound the disc, not of the expansion. For the other models $\varrho$ is not known in closed form, and the series is used only well inside its disc.

\subsection{Continuation of the breather}\label{sec:continuation}

We evaluate the series at a value of $\xi$ inside its disc of convergence to obtain a starting state, and integrate \eqref{eq:galerkin} numerically from there. The details of the numerical method are given in appendix~\ref{sec:numerics}.

\begin{figure}
  \centering
  \includegraphics[width=\textwidth]{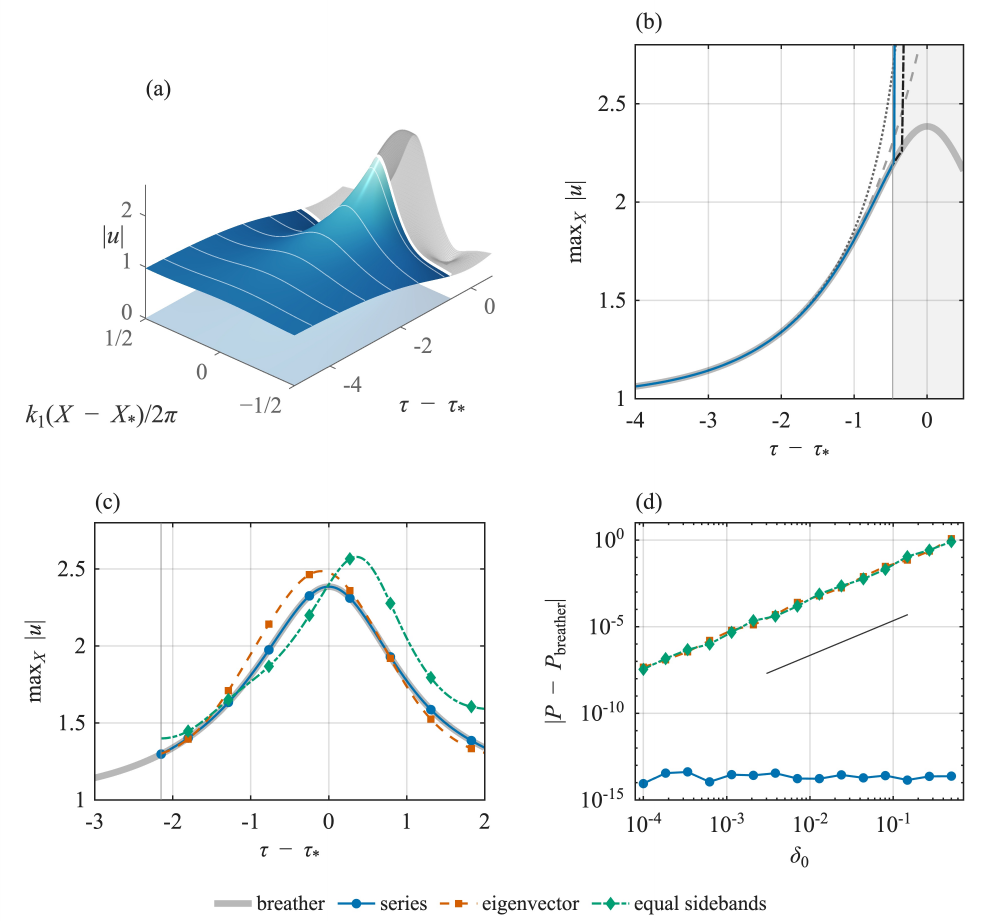}
  \caption{The series and the continuation, for the Dysthe equation with the exact dispersion at $\eps=0.1$ and $\kone=1.3$ ($\Nw=5.4$), $N=32$. (\textit{a}) $\abs{u}$ from the series alone, with $J=240$ terms, up to the instant where it has converged to $10^{-8}$ (white line), and the integrated breather beyond it (grey mesh), over one modulation period centred on the peak. (\textit{b}) $\max_X\abs{u}$ of the partial sums with $J=1$ (dashed), $4$ (dotted), $16$ (dash-dotted) and $240$ (solid), against the breather (thick grey line); the shaded region, beyond the vertical line, lies outside the disc of convergence, $\abs{\xi}>\varrho$. (\textit{c}) Integrations started at $\delta_0=0.2$ (vertical line) from the series (circles), from the unstable eigenvector (squares) and from equal sidebands (diamonds), against the breather (thick grey line). (\textit{d}) The error of the computed peak against the amplitude $\delta_0$ of the start, for the same starts; the thin line has slope $2$.}
  \label{fig:FS}
\end{figure}

Figure~\ref{fig:FS} shows how far the series reaches, and why the starting point matters, for the Dysthe equation with the exact dispersion at $\eps=0.1$ and $\kone=1.3$. The series alone describes the breather far from the Stokes wave, up to about $90\%$ of its peak (panels (\textit{a}) and (\textit{b})). Unlike that of the Akhmediev breather, its disc of convergence, estimated from the growth of the coefficients, ends before the peak. An integration started off the unstable manifold, from the unstable eigenvector or from a pair of sidebands, carries its initial error through the whole growth (panel (\textit{c})). Started at $\delta_0=0.2$, equal sidebands overshoot the peak by more than the whole difference between the peaks of the NLS and Dysthe equations, and the eigenvector by about three quarters as much; the error decreases only like $\delta_0^2$ (panel (\textit{d})). Started from the series, the computed peak does not depend on $\delta_0$ at all, to round-off.

\subsubsection{Truncation errors in the Akhmediev breather}

The breather is computed in two stages: the evaluation of a convergent power series, which represents the exact breather inside its disc of convergence, and a numerical integration beyond it, which requires a truncated model. Here we assess the truncation error in the computation of the Akhmediev breather.

The peak of the Akhmediev breather is exactly
\begin{align}\label{eq:Peak_AB}
\Peak_{AB} = 1 + 2\sin\theta_1,
\end{align}
with $\theta_1 = \arccos(k_1/2)$. Along the evolution the wave action in the carrier is
\begin{equation}\label{eq:abPi}
  \Pi_{AB}(\tau)=1-4\cos^2\thone\Bigl[\frac{\cosh\lamone\tau}{(\cosh^2\lamone\tau-\sin^2\thone)^{1/2}}-1\Bigr],
\end{equation}
its minimum being $\Pi_{AB}(0) = (2\cos\theta_1 - 1)^2$, at the peak.

We can measure truncation errors in terms of the wave action, which is conserved in all models, and in terms of the peak amplitude.

At the peak, $\tau=0$ or $\xi=\varrho$, \eqref{eq:abmodesxi} gives $F=2\cos\thone\,\ee{2\ii\thone}$ and $q=\tan(\thone/2)$. Every Fourier mode then has the same phase $2\thone$, the carrier included, $A_0=(2\cos\thone-1)\ee{2\ii\thone}$, and their moduli are
\begin{equation}\label{eq:abspectrum}
  \abs{A_0}=2\cos\thone-1,\qquad \abs{A_n}=2\cos\thone\,q_1^{\abs{n}}\quad(n\ne0),\qquad
  q_1=\tan\tfrac{\thone}{2}=\Bigl(\frac{2-\kone}{2+\kone}\Bigr)^{1/2}.
\end{equation}
Being in phase, the modes add up at $X=0$, and their moduli sum to the peak \eqref{eq:Peak_AB}. Along the breather consecutive modes are in the ratio $q$, which grows from $\xi\sec\thone$ at the onset of growth, the cascade \eqref{eq:cascade}, to its largest value $q_1$ at the peak.

The amount of wave action that is left out due to truncation is
\begin{equation}\label{eq:mutail}
  \mu_N=\sum_{\abs{n}>N}\abs{A_n}^2=\frac{8\cos^2\thone\;q_1^{2N+2}}{1-q_1^2},
\end{equation}
which decays geometrically, with ratio $q_1^2$.

Let $\Peak_N$ be the peak \eqref{eq:firstevent} computed with the NLS truncated at $N$. The observed error behaves as
\begin{equation}\label{eq:convergence}
  0<1+2\sin\thone-\Peak_N\sim q_1^{\,N}\propto\mu_N^{1/2}.
\end{equation}

\begin{figure}
  \centering
  \includegraphics[width=\textwidth]{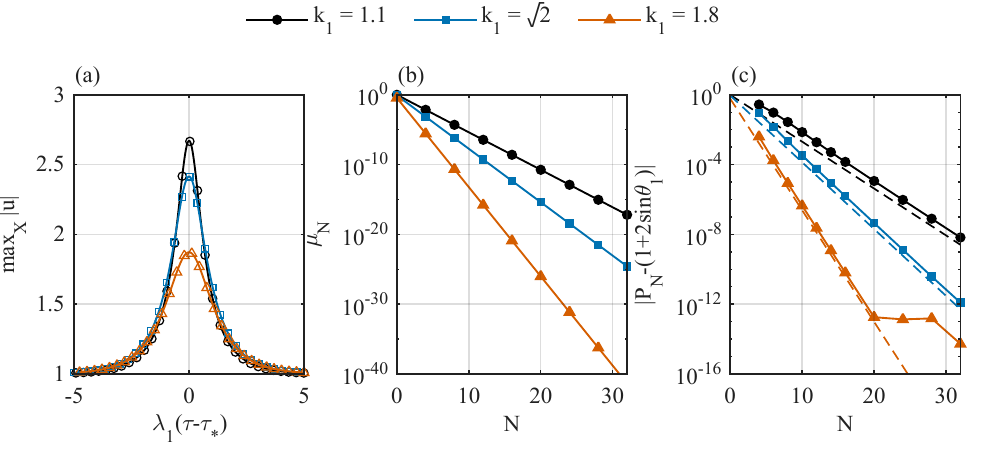}
  \caption{The construction against the exact breather, for the NLS, at $\kone=1.1$, $\sqrt2$ and $1.8$, each with one colour and one marker. (\textit{a}) $\max_X\abs{u}$ through the focusing event against $\lambda_1(\tau-\tau_*)$, computed with $N=32$ (solid lines), and the Akhmediev breather \eqref{eq:abchin} (open markers). (\textit{b}) The wave action $\mu_N$ that the breather carries outside $\abs{n}\le N$ at its peak, \eqref{eq:mutail}, which falls at the geometric rate $q_1^2$, $q_1=\tan(\thone/2)$. (\textit{c}) The error of the computed peak, $\abs{\Peak_N-(1+2\sin\thone)}$, against $N$ (solid lines with filled markers), and $\mu_N^{1/2}$ (dashed lines); at $\kone=1.8$ the error levels off near $10^{-13}$, the accuracy of the integration. The series is taken to order $16$.}
  \label{fig:FD}
\end{figure}

Figure \ref{fig:FD} shows the computed $\max_X\abs{u}$, which follows the breather through the focusing event (panel (\textit{a})), the wave action left out by the truncation, $\mu_N$, which falls geometrically (panel (\textit{b})), and the error of the peak, $\abs{\Peak_N-(1+2\sin\thone)}$, which falls as $\mu_N^{1/2}$ until it reaches the accuracy of the integration (panel (\textit{c})).

Across the band, with $N=32$, the computed peak agrees with $1+2\sin\thone$ to $2\times10^{-8}$, while the computed series coefficients reproduce the closed forms \eqref{eq:a2}, \eqref{eq:a3} and \eqref{eq:diagonal}, and the wave action is conserved, to round-off.

At the peak, where all of the Fourier modes are in phase, their contribution to $\Peak$ is far larger than their share of the wave action suggests.  Take the band $\abs{n}<\Nw$ at $\eps=0.1$ and $\kone=1.3$, that is, $\Nw=5.4$ waves per modulation period. The modes outside it hold only $0.04\%$ of the wave action at the peak, yet, being in phase at the crest, they supply $3\%$ of its height above the background.

\section{Implications for deep water waves}\label{sec:deepwater}

Breathers matter for water waves because they are candidates for \emph{freak waves}: large-amplitude waves that grow out of `nowhere' from an orderly but unstable wave train, reaching several times the amplitude of the surrounding waves. The peak $\Peak$, sometimes called the amplification factor, is the main quantity we study, since it measures how much the waves of the train grow. We now show how it depends on the two physical parameters of the wave train, its steepness $\eps$ and either $\kone$ or $\Nw$, and explain its behaviour.

\subsection{Breathers of the four models}\label{sec:fourmodels}

We first compare the breathers of the four models of table~\ref{tab:models} at $\eps=0.1$.

\begin{figure}
  \centering
  \includegraphics[width=\textwidth]{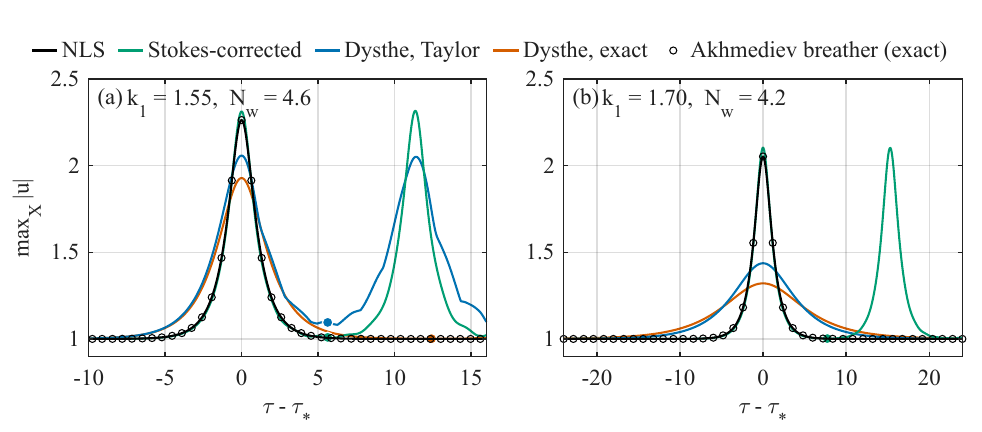}
  \caption{The first peak of the breather for the four models of table~\ref{tab:models} at $\eps=0.10$:
    $\max_X\abs{u}$ against $\tau-\tau_*$ at (\textit{a}) $k_1=1.55$ ($\Nw=4.56$) and (\textit{b}) $k_1=1.7$ ($\Nw=4.16$). Circles: the exact Akhmediev breather. A dot marks $\tau_{\mathrm{ret}}$, the end of the first peak in the sense of \eqref{eq:firstevent}, where it falls in the window. $N=24$ for the reflection-symmetric models and $32$ for the Dysthe equation, $J=16$; with the exact dispersion the modes are $-32\le n<\Nw$.}
  \label{fig:FW}
\end{figure}

\begin{table}
  \centering
  \begin{tabular}{@{}lcclccl@{}}
    \toprule
    & \multicolumn{3}{c}{$\Nw=4.56$ ($k_1=1.55$)} & \multicolumn{3}{c}{$\Nw=4.16$ ($k_1=1.70$)} \\
    \cmidrule(lr){2-4}\cmidrule(l){5-7}
    model & $\Peak$ & $\lambda_1$ & after the peak & $\Peak$ & $\lambda_1$ & after the peak \\
    \midrule
    NLS & $2.264$ & $0.980$ & returns & $2.054$ & $0.896$ & returns \\
    Stokes-corrected & $2.314$ & $1.004$ & repeats & $2.105$ & $0.927$ & repeats \\
    Dysthe & $2.059$ & $0.649$ & repeats & $1.437$ & $0.305$ & returns \\
    Dysthe, exact dispersion & $1.929$ & $0.636$ & returns & $1.321$ & $0.235$ & returns \\
    \bottomrule
  \end{tabular}
  \caption{The first focusing event of figure~\ref{fig:FW} in numbers, for the models of
    table~\ref{tab:models} at $\eps=0.10$: the height $\Peak$ of \eqref{eq:firstevent} and the growth rate
    $\lambda_1$ of the unstable pair. After the first peak the orbit either returns to the Stokes wave, to within
    $3\times10^{-6}$ in $\Pi$ ($10^{-12}$ for the NLS, whose orbit is
    homoclinic), or focuses again: the carrier turns back short of its initial share, by
    $4\times10^{-6}$ to $3\times10^{-3}$, and a second peak of nearly the same height follows. Same
    computations as figure~\ref{fig:FW}.}
  \label{tab:FW}
\end{table}

Figure~\ref{fig:FW} follows the first peak of the four models at two points of the instability band, $\Nw=4.56$ in panel (\textit{a}), $\Nw=4.16$ in panel (\textit{b}), and table~\ref{tab:FW} gives their height and growth rate. For the NLS the computed curve is the Akhmediev breather, drawn over it as circles.

The models differ first in height. The Stokes correction raises the peak slightly, by about $2\%$. The Dysthe equations lower it, by about $10\%$ to $15\%$ at $\Nw=4.56$ and by about a third at $\Nw=4.16$. Between these two nearby points of the instability band their peak falls sharply.

Next, they differ in what follows the peak. The Stokes-corrected orbit does not return to the Stokes wave: a second peak of nearly the same height follows. The Dysthe equation with the Taylor dispersion repeats at $\Nw=4.56$ and returns at $\Nw=4.16$, and the one with the exact multiplier returns to the Stokes wave at both points. At $\Nw=4.16$, where both Dysthe orbits return, their curves of $\max_X\abs{u}$ are symmetric in time, like that of the Akhmediev breather.

We point out that the definition \eqref{eq:firstevent} needs no distinction between these cases. The first peak ends at the first turn of $\Pi$, marked by a filled dot in figure \ref{fig:FW} where it falls in the window.

\subsubsection{The breather of the Dysthe equation with the exact dispersion}\label{sec:dysthebreather}

\begin{figure}
  \centering
  \includegraphics[width=\textwidth]{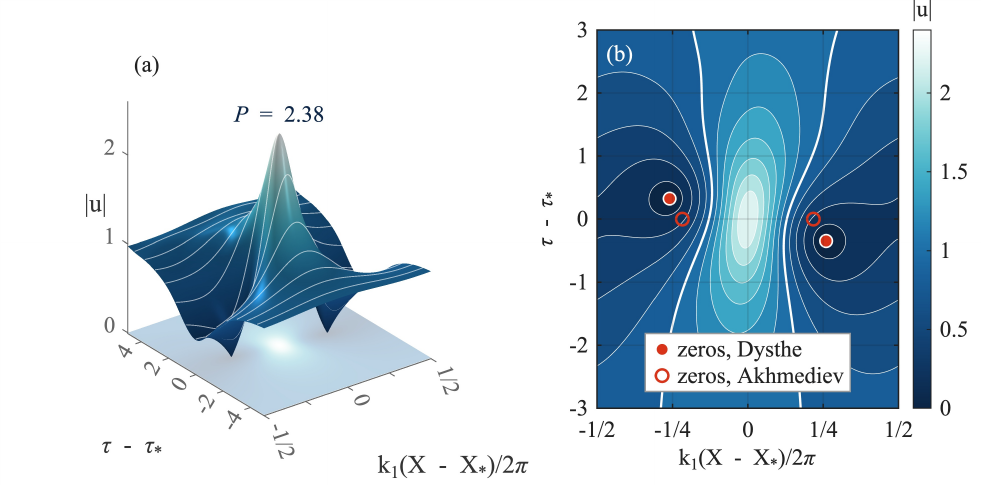}
  \caption{The breather of the Dysthe equation with the exact dispersion, at $\eps=0.10$ and $\kone=1.3$ ($\Nw=5.4$), over one modulation period centred on the peak. (\textit{a}) $\abs{u(X,\tau)}$ in the drift frame, time running into the page; the colour follows the height, from the blue of the undisturbed train ($\abs{u}=1$) to white at the crest, the white lines are the envelope at fixed $\tau$, every $0.75$, and the floor carries the same field seen from above. (\textit{b}) The same field as a contour map, with the level $\abs{u}=1$ in bold. Filled circles, the zeros of $\abs{u}$; open circles, the zeros of the Akhmediev breather at the same $\kone$. $N=32$ (modes $-32\le n\le5$), $J=16$. }
  \label{fig:FWD}
\end{figure}

Figure~\ref{fig:FWD} shows the breather of the Dysthe equation with the exact dispersion at $\eps=0.1$ and $\Nw=5.4$, computed as in section~\ref{sec:construction}. It is the analogue of the Akhmediev breather in deep water. It differs from it in two main ways. First, its peak is about $5\%$ lower than that of the Akhmediev breather at the same $\kone$. Second, it has a visible temporal asymmetry around the peak, which the contour map of figure~\ref{fig:FWD}(\textit{b}) shows best. There $\abs{u}$ vanishes at two points per modulation period. In the Akhmediev breather both zeros occur at the instant of the peak, symmetrically on either side of it. In the Dysthe breather they lie slightly farther from the peak and are staggered in time. The zero ahead of the peak (larger $X$) occurs before it, and the one behind occurs after it.

We regard the Dysthe equation as a more complete model of deep water wave trains than the NLS, with or without the Stokes correction, and of its two versions we take the one with the exact dispersion as our deep-water model. The reason, given in section~\ref{sec:threemodels}, is the group velocity: the Taylor dispersion makes the short waves that a focusing group excites travel faster than the group, while in deep water they are slower. From here on we work with the Dysthe equation with the exact dispersion.

\subsection{The map of the envelope amplification}\label{sec:map}

\begin{figure}
  \centering
  \includegraphics[width=\textwidth]{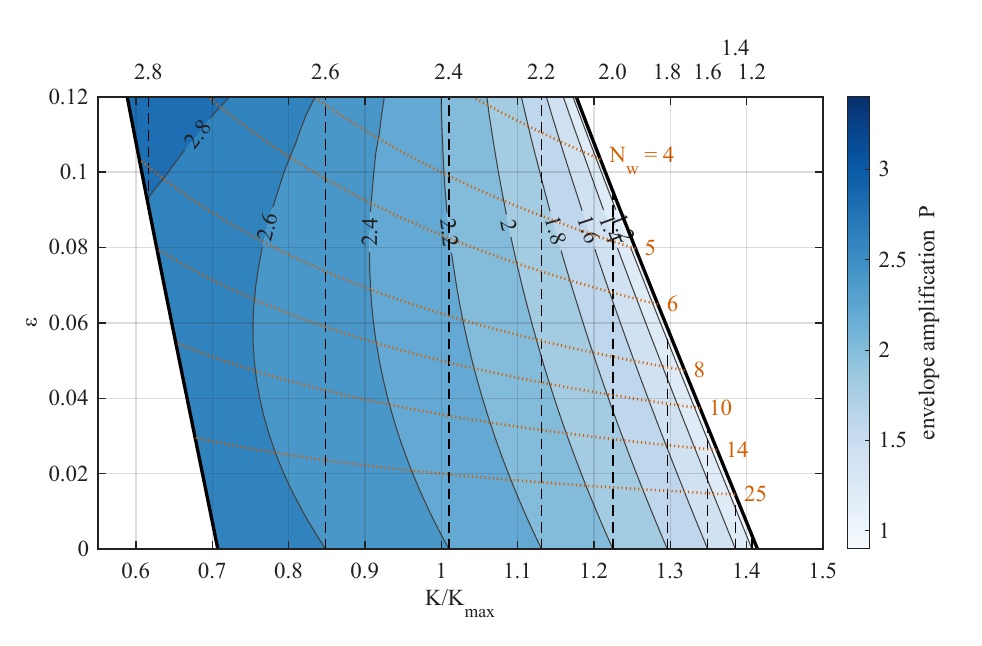}
  \caption{Envelope amplification $\Peak$ of the first peak over $\Kmod/\Kmod_{\max}$ and $\eps$ for the Dysthe equation with the exact dispersion. The map covers the instability band of the model at each $\eps$, the thick lines are its edges. The dashed lines are the NLS values $1+2\sin\thone$ at the same levels, which do not depend on $\eps$; each value is printed above its line. The row $\eps=0$ is that limit, to which the Dysthe equations reduce. The dotted curves are lines of constant $\Nw=1/(2\eps\,\Kmod/\Kmod_{\max})$. The model is truncated at $N=32$, so the modes are $-32\le n<\Nw$.}
  \label{fig:FA}
\end{figure}

Our main result for deep water is the map of figure~\ref{fig:FA}, which gives $\Peak$ for the Dysthe equation over the steepness and the modulation wavenumber, the latter scaled by the most unstable one of the NLS, $\Kmod/\Kmod_{\max}=\kone/\sqrt2$. On that abscissa the NLS value $1+2\sin\thone$ does not depend on $\eps$, so its contours are vertical lines, drawn dashed. The Dysthe contours coincide with them as $\eps\to0$ and bend away as the steepness grows.

Over most of the map $\Peak$ lies below the value of the NLS.  Values at selected points are given in table~\ref{tab:T1}. For the longest groups the two are within a few per cent, but the deficit grows towards the short-group edge of the band, where it reaches $20\%$ to $30\%$. The Dysthe peak rises above the NLS one only for the longest groups of the steepest wave trains, from $\eps\approx0.105$ on: at $\eps=0.12$ and $\Nw=5.8$ it is $3\%$ higher.

Each peak computed on the map has a single hump and the contours are smooth. This is not so with the Taylor dispersion, whose map is qualitatively similar but less regular in its interior. Table~\ref{tab:T1}, which gives its values at the same points, shows peaks with two humps, marked with a dagger.

Table \ref{tab:T1} collects the four models at seven points of their common instability band at $\eps=0.05$ and $0.10$, and at the point at $\eps=0.12$ discussed above. It also gives the linear prediction $1 + 2\sin\theta_1$, with $\theta_1$ given by equation \eqref{eq:theta} for the Dysthe equation with the exact dispersion, and the number of carrier periods in which the instability grows by a factor $\ee{}$, for that equation and, in brackets, for the NLS. This number follows from the linear stability analysis and the carrier period. The unstable mode grows like $\ee{\lamone\tau}$, so it grows by a factor $\ee{}$ in $\Delta\tau=1/\lamone$, which by \eqref{eq:vars} is the physical time $2/(\eps^2\wc\lamone)$; divided by the carrier period $2\pi/\wc$, it gives $1/(\pi\eps^2\lamone)$ carrier periods. It is a linear estimate, but a useful one: a modulation imposed with a small relative amplitude $\delta$ needs about $\ln(1/\delta)$ such growth times to focus, which sets the duration of an experiment or, since the group travels half a carrier wavelength per period, the length of the tank.

\begin{table}
  \centering
  \setlength{\tabcolsep}{5pt}
  \begin{tabular}{@{}cccccccc@{}}
    \toprule
    & & \multicolumn{4}{c}{envelope amplification $\Peak$} & & periods \\
    \cmidrule(lr){3-6}
    $\eps$ & $\Nw$ & NLS & Stokes-corrected & Dysthe & Dysthe, exact & $1+2\sin\thone$ & per $\ee{}$-fold \\
    \midrule
    $0.05$ & $14.0$ & $2.726$ & $2.741$ & $2.690$ & $2.641$ & $2.703$ & $154$ ($146$) \\
     & $12.0$ & $2.616$ & $2.630$ & $2.552$ & $2.512$ & $2.575$ & $143$ ($134$) \\
     & $10.0$ & $2.414$ & $2.427$ & $2.300$ & $2.273$ & $2.331$ & $143$ ($127$) \\
     & $9.0$ & $2.237$ & $2.250$ & $2.073$ & $2.052$ & $2.100$ & $156$ ($131$) \\
     & $8.4$ & $2.080$ & $2.092$ & $1.856$ & $1.838$ & $1.875$ & $184$ ($140$) \\
     & $8.0$ & $1.935$ & $1.949$ & $1.632$ & $1.612$ & $1.638$ & $241$ ($154$) \\
     & $7.8$ & $1.844$ & $1.858$ & $1.460$ & $1.436$ & $1.453$ & $332$ ($166$) \\
    \addlinespace
    $0.10$ & $7.0$ & $2.726$ & $2.785$ & $2.579^{\dagger}$ & $2.713$ & $2.675$ & $41$ ($37$) \\
     & $6.4$ & $2.667$ & $2.724$ & $2.553^{\dagger}$ & $2.619$ & $2.595$ & $39$ ($35$) \\
     & $5.5$ & $2.532$ & $2.585$ & $2.449$ & $2.405$ & $2.399$ & $39$ ($32$) \\
     & $4.8$ & $2.353$ & $2.403$ & $2.191$ & $2.100$ & $2.102$ & $44$ ($32$) \\
     & $4.6$ & $2.279$ & $2.329$ & $2.085$ & $1.960$ & $1.962$ & $49$ ($32$) \\
     & $4.4$ & $2.191$ & $2.240$ & $1.865$ & $1.767$ & $1.766$ & $59$ ($33$) \\
     & $4.2$ & $2.080$ & $2.130$ & $1.536$ & $1.439$ & $1.435$ & $100$ ($35$) \\
    \addlinespace
    $0.12$ & $5.8$ & $2.723$ & $2.807$ & $2.418^{\dagger}$ & $2.807$ & $2.657$ & $29$ ($25$) \\
    \bottomrule
  \end{tabular}
  \caption{Envelope amplification $\Peak$ of the first peak for the four models, at seven values of $\Nw$ for $\eps=0.05$ and $0.10$, and at one point at $\eps=0.12$ where the Dysthe equation with the exact dispersion rises above the NLS; every $\kone=1/(\sqrt2\eps\Nw)$ lies in the instability band of all four. The last three rows at $\eps=0.05$ lie in the short-group fifth of the band of the Dysthe equation with the exact dispersion, the last four at $\eps=0.10$ in its short-group third. The column $1+2\sin\thone$ is the linear prediction of section~\ref{sec:decomposition}, the Akhmediev law evaluated with the angle $\thone$ of the Dysthe equation with the exact dispersion. The last column is the number of carrier periods, $1/(\pi\eps^2\lamone)$, in which the instability grows by a factor $\ee{}$ in that equation, with the NLS value in brackets. A dagger marks a focusing event with two humps, of which $\Peak$ is the higher. Every entry is converged in the truncation to $10^{-3}$ or better, the entry of the Dysthe equation with the exact dispersion at $\eps=0.12$ being computed with $N=48$.}
  \label{tab:T1}
\end{table}

\subsection{A linear summary of the map}\label{sec:decomposition}

The map is summarised, to a good approximation, by the \emph{Akhmediev law} evaluated with the linear data of the Dysthe equation with the exact dispersion,
\begin{equation}\label{eq:linlaw}
  \Peak\approx1+2\sin\thone,\qquad
  \cos2\thone=\frac{\Omega_{\mathrm e}(k_1)-1+1/\Nw}{\bigl[(1-\tfrac{3}{2\Nw})(1-\tfrac{1}{2\Nw})\bigr]^{1/2}},
\end{equation}
where $\Omega_{\mathrm e}(k)=\tfrac12[\Omega(k)+\Omega(-k)]$ is the even part of the exact multiplier \eqref{eq:exactdisp}. The angle is that of \eqref{eq:theta}, in the drift frame $\alpha_1=\alpha_{-1}=\Omega_{\mathrm e}(k_1)-1+1/\Nw$ and $\beta_{\pm1}=1-(2\pm1)/(2\Nw)$, by \eqref{eq:dysthelin}. The prediction needs only the linear stability of the Stokes wave, through the angle of its unstable eigenvector, and no nonlinear computation. That it holds is, however, a result of the nonlinear computation: it is what the first peaks of the breathers, computed across the map of figure~\ref{fig:FA}, show.

For every steepness up to $\eps=0.1$, and everywhere in the instability band, $\Peak$ and $1+2\sin\thone$ differ by less than $3\%$; table~\ref{tab:T1} gives them side by side. The agreement is best for short groups and degrades towards the long-group edge of the band, more so for steeper waves: at $\eps=0.12$ the difference there reaches $6.5\%$.

The law also accounts for the shape of the map. As $k_1$ approaches the short-group edge of the band the growth rate $\lamone$ vanishes, and with it $\thone$, by \eqref{eq:theta}, so that $\Peak$ falls to one: the steep fall of the amplification towards short groups in figure~\ref{fig:FA} is the approach to the edge of the instability band of the Dysthe equation. In the Dysthe model the height of the peak is thus summarised, to within a few per cent, by a single number from the linear stability of the Stokes wave, the angle $\thone$ of its unstable eigenvector.

\subsection{Wave height}\label{sec:threequantities}

A wave gauge records the surface, not the envelope, but it can be reconstructed from the envelope with the help of \eqref{eq:recon}. There are three quantities that describe the maximum wave elevation that we explore herein. These are the maximum envelope amplification $\Peak$ of the breather together with the crest and the wave height amplification factors which are obtained from the surface wave. 

To this end,  wave crest, trough and wave height must be defined. Individual waves are defined by the zero-crossing convention of oceanography: the record of the surface is divided into \emph{waves} at its zero up-crossings, the points where it crosses the mean water level upwards, which is one of the two usual choices \citep[section~3.3]{holthuijsen2007}. The crest and trough elevations are the maximum and the minimum of each wave, and the height of a wave is its crest elevation minus its trough elevation.  The crest amplification factor $P_{\mathrm{crest}}$ is the largest crest in the whole wave record divided by the crest of the undisturbed wave train, $a_0(1+\eps/2+3\eps^2/8)$ by \eqref{eq:recon}, and the height amplification factor $P_{\mathrm{H}}$ is the largest wave height divided by the height of the undisturbed wave train, $2a_0(1+3\eps^2/8)$.

Unlike the envelope amplification $\Peak$, which depends only on the envelope, the largest crest and the largest wave height on the surface at a given instant also depend on the phase of the carrier relative to the group. The envelope equations leave this phase free, since they are unchanged when $u$ is multiplied by a constant phase factor.  The carrier does not, however, keep its phase relative to the group. Its crests travel at the phase speed, twice the group speed, so they move through the group, each growing and decaying as it passes \citep[p. 121]{banner_tian1998}. By \eqref{eq:vars} the carrier phase in \eqref{eq:recon} is $\chi=\tau/\eps^2-X/(\sqrt2\eps)+\chi_0$, with a constant $\chi_0$, so at a fixed $X$ it advances by $\pi$ in each carrier period, $\pi\eps^2$ in $\tau$, while $u$ evolves on the scale $\tau=O(1)$.  At every point of table~\ref{tab:T1} the maximum of $\abs{u}$ stays within $1\%$ of $\Peak$ for at least four carrier periods.

Every carrier phase is therefore realised while the envelope is at its peak, and the largest crest and the largest wave height are attained at $\tau_*$, the instant of the peak. We take $P_{\mathrm{crest}}$ and $P_{\mathrm{H}}$ to be these maxima.  Like $\Peak$, these two factors are thus properties of the breather and not of its carrier phase, as \citet[p.~244]{slunyaev_shrira2013} argued for envelopes that evolve slowly compared with the wave period.

Reconstructing the surface at every instant of that time, for $16$ values of $\chi_0$, confirms this: for every $\chi_0$ the largest crest and the largest wave height agree with the maxima at $\tau_*$ to within $0.3\%$.

\begin{figure}
  \centering
  \includegraphics[width=\textwidth]{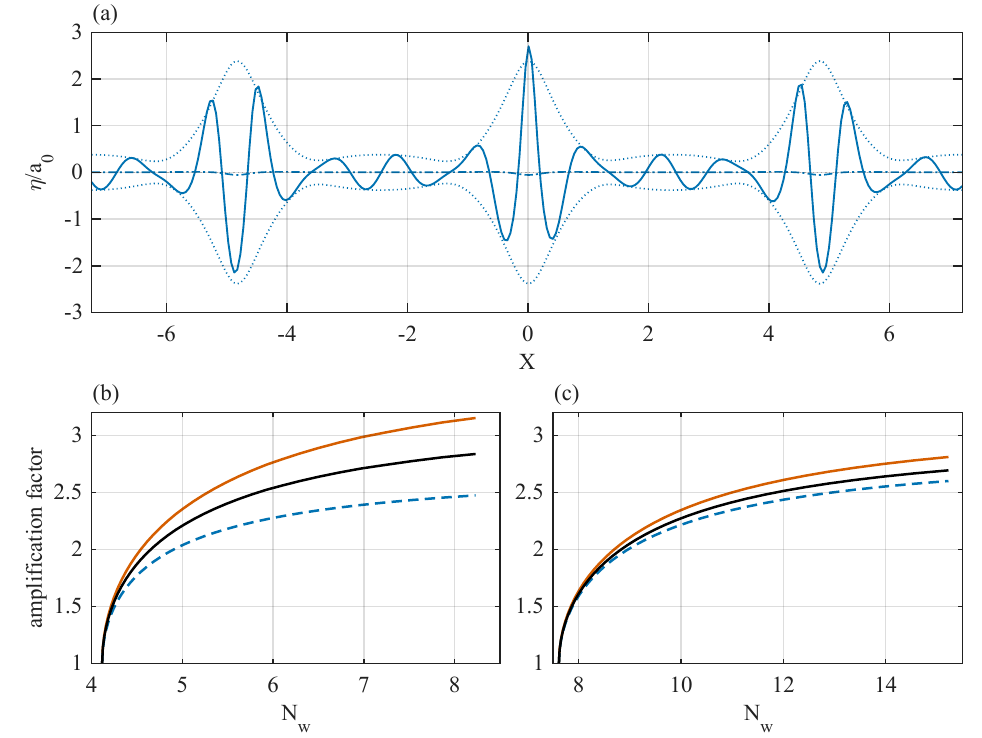}
  \caption{(\textit{a}) The surface reconstructed with \eqref{eq:recon} at the instant of the peak, at $\eps=0.1$ and $\kone=1.3$ ($\Nw=5.4$), for the Dysthe equation with the exact dispersion, at its crest-maximising carrier phase, over three modulation periods centred on the peak. Solid, $\eta/a_0$; dotted, $\pm\abs{u}$; dash-dot, the induced mean level. (\textit{b}), (\textit{c}) The amplification factors of the same equation, against $\Nw$ over the instability band, at (\textit{b}) $\eps=0.10$ and (\textit{c}) $\eps=0.05$: the envelope amplification $\Peak$ (black), the crest amplification factor $P_{\mathrm{crest}}$ (orange, solid) and the height amplification factor $P_{\mathrm{H}}$ (blue, dashed). All three fall to one at the short-group edge of the band. Same computations as figure~\ref{fig:FA}, with $256$ carrier phases.}
  \label{fig:FC}
\end{figure}

Figure \ref{fig:FC}(\textit{a}) shows the surface reconstructed with \eqref{eq:recon} at the peak, for the exact-dispersion Dysthe equation at $\eps=0.1$ and $\kone=1.3$. The bound harmonics raise the crest above the envelope maximum, because the second harmonic adds to a crest and subtracts from a trough.

Figure~\ref{fig:FC}(\textit{b}) and (\textit{c}) show $P_{\mathrm{crest}}$ and $P_{\mathrm{H}}$ across the instability band at the two steepnesses. The two factors lie on either side of the envelope amplification everywhere in the band: the largest crest exceeds $\Peak$ and the largest wave height falls below it, by up to $10\%$ and $12\%$, respectively, at $\eps=0.1$. Both differences grow with the steepness and with $\Peak$.

Because the reconstruction \eqref{eq:recon} is additive, the contribution of each of its terms can be found by switching the terms on one after another, the largest crest and height being found anew each time. The free wave alone puts its largest crest at the peak of the envelope, where it is $\Peak a_0$. The bound harmonics raise it, by up to about $20\%$ at $\eps=0.1$, and the mean level lowers it by a few per cent. The bound harmonics raise the crest by more than those of a uniform Stokes wave of amplitude $\Peak$ would, and the mean level, which a uniform wave does not have, takes this excess away again almost exactly. The largest crest is therefore that of a local Stokes wave of amplitude $\Peak$. For a uniform envelope $u=\Peak$ the mean level and the derivative term in \eqref{eq:recon} vanish, and the free wave and its two harmonics peak together at $\chi=0$, so the crest is $a_0(\Peak+\eps\Peak^2/2+3\eps^2\Peak^3/8)$. Dividing it by the crest of the undisturbed train, the same expression with $\Peak=1$, gives
\begin{equation}\label{eq:crestlaw}
  P_{\mathrm{crest}}\approx\frac{\Peak+\eps\Peak^2/2+3\eps^2\Peak^3/8}{1+\eps/2+3\eps^2/8},
\end{equation}
which has no adjustable constants and holds to within $1\%$ over the whole range of figure~\ref{fig:FA}.

The height behaves differently. The crest and the trough of a wave lie half a carrier wavelength apart, so they cannot both sit at the peak of the envelope, and the largest height of the free wave alone falls short of $2\Peak a_0$, by up to about $15\%$ at $\eps=0.1$. The second harmonic raises a crest and a trough alike, so the bound harmonics restore only a few per cent of this, and the mean level changes it even less.

\subsection{Comparison with fully nonlinear breathers}\label{sec:euler}

The breathers of the Dysthe equation stand for breathers of the full water-wave equations, and these have been computed. \citet{slunyaev_shrira2013} perturbed a Stokes wave train along a single unstable mode of the NLS and followed it with simulations of the potential Euler equations, for steepnesses $s_0$ (we are keeping their notation) from $0.04$ to $0.15$ and from $3$ to $20$ waves per modulation. Their $s_0$ is the steepness of the free wave of the undisturbed train and their $\Nw$ the number of carrier waves in one modulation, our $\eps$ and $\Nw$. For each simulation they recorded the highest crest and the largest height, the highest crest minus the deepest trough, reached during the first focusing, each divided by its value at the start \citep[eqs.~4.1 and~4.3]{slunyaev_shrira2013}.

They found that the NLS describes these breathers only qualitatively.  Modulations that it predicts to grow do not grow at all, and the crests grow higher than its envelope. To describe their simulations they introduced two fits: a shift of $0.47$ in the number of waves of the instability criterion \citep[eq.~4.4]{slunyaev_shrira2013}, and an empirical increase of the envelope amplification, with two fitted constants, before dressing the envelope as a local Stokes wave \citep[eqs.~4.6 and~4.7]{slunyaev_shrira2013}. The result reproduces their crests and troughs with a root-mean-square relative error of up to $15\%$ \citep[p.~229]{slunyaev_shrira2013}.

We read their simulations off their figure~6 and computed our breather at those that lie in the band of a single unstable pair of the Dysthe equation with the exact dispersion, with $\eps=s_0\le0.125$: $52$ simulations, with $\Nw$ from $4$ to $15$. Their simulations start from Stokes waves of steepness $s=k_0H/2$, where $H$ is the wave height, bound harmonics included, while their figures are drawn against $s_0$. For a Stokes wave $s=s_0(1+3s_0^2/8)$, so we take $\eps=s_0=s/(1+3s^2/8)$ at their values of $s$, in agreement with the correspondence they quote \citep[p.~222]{slunyaev_shrira2013}.

Two further points need care. Their indices are maxima over the whole first focusing, while ours are taken at $\tau_*$, which by section~\ref{sec:threequantities} gives the same values; and at all $52$ points the highest crest and the deepest trough belong to the same zero-crossing wave, so that their height index is our $P_{\mathrm{H}}$. Their indices are, moreover, divided by the crest and the height of their initial condition, which carries their perturbation, of relative amplitude $0.01$ \citep[p.~213]{slunyaev_shrira2013}, along the eigenvector of the NLS. By \eqref{eq:eigvec} this perturbation is $\ee{\ii\thone}$ times a real modulation, with $\cos\thone=\tfrac12k_1$ for the NLS, so only the fraction $\tfrac12k_1$ of it, the part in phase with the carrier, changes $\abs{u}$ (this fraction is $1/R_{\mathrm{BF}}$ in their notation, \citealp[eq.~2.5]{slunyaev_shrira2013}). The initial envelope maximum is therefore $1+0.005\,k_1$, and we divide our factors by the same amount.

\begin{table}
  \centering
  \setlength{\tabcolsep}{8pt}
  \begin{tabular}{@{}ccccc@{}}
    \toprule
    & & & \multicolumn{2}{c}{Euler} \\
    \cmidrule(l){4-5}
    $\Nw$ & NLS & Dysthe, exact & fit & first growing simulation \\
    \midrule
    $3$ & $0.1179$ & $0.1482$ & $0.1397$ & $0.1389$ ($1.04$) \\
    $4$ & $0.0884$ & $0.1035$ & $0.1002$ & $0.1015$ ($1.29$) \\
    $5$ & $0.0707$ & $0.0797$ & $0.0780$ & $0.0798$ ($1.37$) \\
    $6$ & $0.0589$ & $0.0649$ & $0.0639$ & $0.0649$ ($1.28$) \\
    $7$ & $0.0505$ & $0.0548$ & $0.0541$ & $0.0549$ ($1.28$) \\
    $8$ & $0.0442$ & $0.0474$ & $0.0470$ & $0.0469$ ($1.03$) \\
    $9$ & $0.0393$ & $0.0418$ & $0.0414$ & $0.0420$ ($1.26$) \\
    \bottomrule
  \end{tabular}
  \caption{The steepness $s_0=\eps$ at which a modulation of $\Nw$ waves begins to grow: for the NLS, $1/(2\sqrt2\Nw)$; for the Dysthe equation with the exact dispersion, the edge of its instability band, $\thone=0$ in \eqref{eq:linlaw}; and for the Euler equations, the fit $1/[2\sqrt2(\Nw-0.47)]$ of \citet[eq.~4.4]{slunyaev_shrira2013} and the first of their simulations that grows, with its crest amplification in brackets, read from their figure 6. }
  \label{tab:T5}
\end{table}

\begin{figure}
  \centering
  \includegraphics[width=\textwidth]{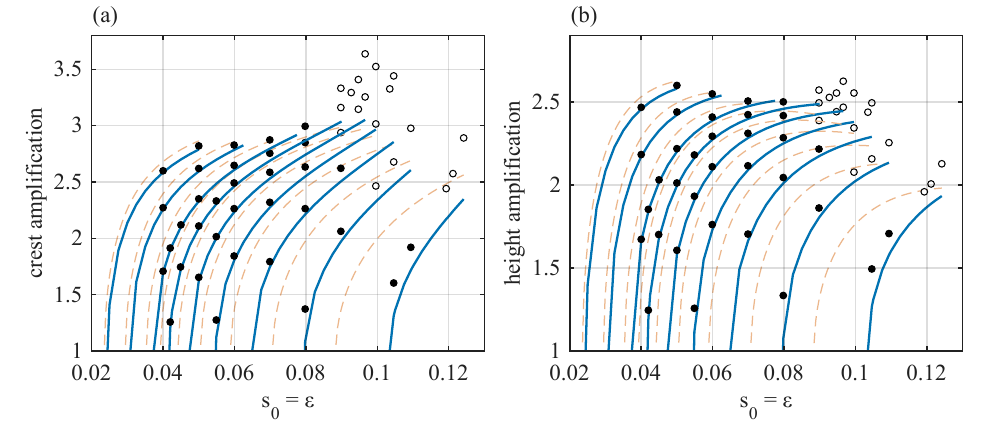}
  \caption{Crest (\textit{a}) and height (\textit{b}) amplification of the fully nonlinear breathers of \citet{slunyaev_shrira2013} (symbols, read from their figure~6) and of the Dysthe equation with the exact dispersion (solid), against the steepness $s_0=\eps$, for $\Nw=15$, $12$, $10$, $9$, $8$, $7$, $6$, $5$ and $4$, from left to right. The height is the highest crest minus the deepest trough, as they define it, which at these points is $P_{\mathrm{H}}$. Only their simulations in the band of a single unstable pair of the Dysthe equation, with $s_0\le0.125$, are shown.  Filled circles correspond to a maximum crest steepness $k_0\eta_{\max}$ below $0.25$ and open circles to one above. Dashed: the NLS breather, its surface reconstructed in the same way. The values of the models are divided by $1+0.005\,\kone$, the normalisation used in the simulations.}
  \label{fig:FE}
\end{figure}

The first comparison is where growth begins. At a given steepness a modulation grows only if the group is long enough, $\Nw>\Nw^*(\eps)$, the edge of the instability band. For the NLS $\Nw^*=1/(2\sqrt2\eps)$. For the Dysthe equation, setting $\thone=0$ in \eqref{eq:linlaw} with $\Omega_{\mathrm e}(\kone)\approx\kone^2/2$ and $\kone^2=1/(2\eps^2\Nw^2)$ gives, to leading order, $\Nw(\Nw-1)=1/(8\eps^2)+1/16$, that is, $\Nw^*\approx1/(2\sqrt2\eps)+1/2$. The half wave comes from the terms in $1/\Nw$ of $\alpha_{\pm1}$ and $\beta_{\pm1}$, which are the terms of order $\eps$ of the Dysthe equation. With the exact dispersion the shift $\Nw^*-1/(2\sqrt2\eps)$ grows from $1/2$ as $\eps\to0$ to about $0.6$ at $\eps=0.1$, while for the Euler equations \citet[eq.~4.4]{slunyaev_shrira2013} fitted a shift of $0.47$ to their simulations. The three edges are therefore ordered, NLS, Euler, Dysthe, with the Dysthe edge much closer to the Euler one.

Table~\ref{tab:T5} reads the same edges the other way, as the steepness above which a modulation of $\Nw$ waves grows, and sets them beside the simulations. At every $\Nw$ the edge of the NLS lies below the Euler fit, and that of the Dysthe equation lies just above it, several times closer. The NLS therefore predicts growth that the Euler equations do not show, while the Dysthe equation misses growth only in a narrow range just above the Euler onset. The first simulation that grows, in the last column, bears this out: it lies on the edge of the Dysthe equation or below it, and where it lies on the edge its amplification is already well above one. Growth in the Euler equations therefore begins slightly below the edge of the Dysthe equation.

Above the onset the amplification itself agrees, figure~\ref{fig:FE}. For $\Nw\ge6$, once the crest amplification exceeds $1.5$ and while the maximum crest steepness $k_0\eta_{\max}$ stays below $0.25$, our crest and height amplifications lie less than $5\%$ below the Euler ones, over $28$ simulations. The NLS breather, reconstructed in the same way and dashed in figure~\ref{fig:FE}, is off by up to about $20\%$, mostly above. The linear data alone do as well: the crest law \eqref{eq:crestlaw} with $1+2\sin\thone$ in place of $\Peak$ gives the Euler crest at the same points to within about $3\%$.

Our breather falls short of the Euler one in two places. Just above the onset, for $\Nw=4$ and $5$, the curves rise steeply, and the small difference in the onset becomes a large one in the amplification, of about $20\%$. And as the crest steepness rises past about $0.25$, towards the steepest crests that \citet{slunyaev_shrira2013} report without breaking, our crests fall below the Euler ones by up to about $20\%$, and our heights by up to about $10\%$. There the crest of the extreme wave approaches that of the limiting Stokes wave and the amplification rises sharply as breaking approaches \citep[p.~222]{slunyaev_shrira2013}, effects of higher order than the Dysthe equation and its reconstructed surface wave.  For the shortest groups, $\Nw=4$ and $5$, the overestimate of the envelope by the NLS partly offsets the missing sharpening there, and its crest comes closer to the Euler one than ours; for longer groups both fall short by similar amounts.

The Dysthe breather therefore accounts, with no fitted constant, for the onset of growth, through the edge of its band, and, below the approach of breaking, for the amplification, through its envelope amplification $\Peak$ and the crest of a local Stokes wave of that amplitude, the two features that \citet{slunyaev_shrira2013} described by fitting the NLS. What it leaves out is the sharpening of the crest as breaking approaches. Up to that point the highest crest of a group grown from a Stokes wave train, which they obtained from the Euler equations, follows from the map of $\Peak$ and, to within a few per cent, from the single angle $\thone$ of the Dysthe equation.

\section{Conclusions}\label{sec:conclusions}

In this paper we have defined and computed the breather of non-integrable envelope models for deep water waves. The starting point is the unstable manifold of the Stokes wave. When a single pair of modes is unstable, gauge and translation invariance are enough to show that this manifold consists of the Stokes wave and of the translates and time shifts of a single solution, the breather. The breather is therefore unique, and its peak is a property of the model and of its parameters, not of an initial condition. Integrability, a Hamiltonian structure, the reflection symmetry and time reversibility play no part, and for the NLS this is the Akhmediev breather.

We computed the breather with the parameterisation method, adapted to a Stokes wave that is not a hyperbolic fixed point. The translations reduce the problem to a single real variable, $\xi=\ee{\lambda_1\tau}$, and the breather becomes a power series in $\xi$ whose coefficients follow recursively from linear systems that can always be solved. We stress again that this series is not an asymptotic expansion. For the NLS it converges, it is the Taylor series of the breather itself, and we found its radius of convergence exactly, $\varrho=\lambda_1/2$ for the eigenvector normalised as in \eqref{eq:a1}, which is set by singularities of the breather in complex time that lie directly above its peak. Hence, the series describes the whole growth of the wave up to the peak. For the Dysthe equations the series converges for every truncation, and we return to the full equations at the end of this section. Beyond the disc of convergence the breather is continued by integrating the equations from a point inside it, which lies on the unstable manifold, and the peak of the Akhmediev breather is recovered to $2\times10^{-8}$. The leading coefficients of the series give the instability cascade in every model: symmetric for the NLS models and asymmetric for the Dysthe equations.

For deep water we used the Dysthe equation with the exact dispersion. We defined the first peak of its breather through the exchange of wave action between the carrier and the sidebands, which does not require the breather to return to the Stokes wave. Our main physical result is the map of the peak amplification factor $\Peak$, over the instability band, for steepness up to $0.12$. The peak lies below that of the NLS, except for the longest groups of the steepest wave trains, and the map is summarised, to within $3\%$ for steepness up to $0.1$, by the Akhmediev formula $1+2\sin\thone$ evaluated with the angle $\thone$ of the unstable eigenvector of the Dysthe equation. The formula is linear and easy to evaluate, but the fact that it holds is a result of the nonlinear computations. On the reconstructed surface, the highest crest is that of a local Stokes wave of amplitude $\Peak$, to within $1\%$ and with no fitted constants, while the height amplification stays below $\Peak$.

The comparison with the Euler simulations of \citet{slunyaev_shrira2013} supports these results. Where growth begins, the edge of the instability band of the Dysthe equation lies just above the Euler onset, while that of the NLS lies well below it. The Dysthe equation therefore misses growth over a range of steepness several times narrower than the range over which the NLS predicts growth that does not happen. Above the onset, for groups of six or more waves and until the crests approach breaking, our crest and height amplifications lie less than $5\%$ below the Euler ones. What \citet{slunyaev_shrira2013} described by fitting the NLS follows here from the Dysthe equation with no fitted constant.

Two things are left out. First, near breaking, the Euler crests sharpen beyond those of the Dysthe equation and its reconstruction, which is an effect of higher order than the model,  and second, we have restricted ourselves to a single unstable pair. With more unstable pairs the unstable manifold is larger and contains combination breathers \citep[section~4]{ablowitz_herbst1990}, but the parameterisation method still applies, provided the unstable eigenvalues are not resonant \citep[definition~3.1]{reinhardt_mj2019}.  Broad-banded models, such as the Zakharov equation, are also gauge and translation invariant, and a three-mode version of their breather has already been found as a heteroclinic orbit \citep{andrade_stuhlmeier2023}. The definition and the construction presented here apply to them as well.

Finally, we computed the breather of the Dysthe equations from their truncations. For every truncation the theorem of \citet{cabre_etal2003} guarantees that the unstable manifold exists and that the series converges. For the full Dysthe equations we do not know whether the series converges, which would also prove that the unstable manifold exists. Our results suggest that it does: as the truncation is refined the breathers converge, and their peaks are converged to $10^{-3}$ or better at every point of table~\ref{tab:T1}. We hope that this will motivate further work on the question.

\backmatterhead{Acknowledgements.}
The author used the AI assistant Claude (Anthropic; models of the Claude family, including Claude Opus~5.5).  It was used to write and check the MATLAB and Python code for the computations and the figures, to digitise figure~6 of \citet{slunyaev_shrira2013} from the vector graphics of the published article, to check the cited statements against the cited sources, and to draft and edit parts of the text. The author directed the work, checked the code, the results and the text, and takes full responsibility for the content of the paper.

\backmatterhead{Funding.}
This research received no specific grant from any funding agency, commercial or not-for-profit sectors.

\backmatterhead{Declaration of interests.}
The author reports no conflict of interest.

\backmatterhead{Data availability statement.}
The data and code that support the findings of this study are available from the author upon request.

\appendix
\makeatletter
\newcommand{\prefix@section}{Appendix~}
\makeatother
\renewcommand{\theequation}{\thesection\,\arabic{equation}}

\section{Numerical details}\label{sec:numerics}

All the computations were done in MATLAB R2025b. Here we give what is needed to reproduce them.

\textbf{The truncated system.} We integrate \eqref{eq:galerkin} in the frame that rotates with the Stokes wave and drifts with the velocity $V$ of \eqref{eq:drift}. The modes kept are $\abs{n}\le N$, and for the Dysthe equation with the exact dispersion $-N\le n\le\min(N,\lceil\Nw\rceil-1)$, as in section~\ref{sec:models}. The nonlinear terms are evaluated with fast Fourier transforms on a grid of $M_{\mathrm{g}}$ points. If the kept modes span $w=n_{\max}-n_{\min}$, the cubic terms are projected back onto them without aliasing when $M_{\mathrm{g}}>2w$, and the quintic term of the Stokes-corrected equation when $M_{\mathrm{g}}>3w$; we take the smallest convenient $M_{\mathrm{g}}$ above these bounds. Unless stated otherwise, $N=32$; the entry of the Dysthe equation with the exact dispersion at $\eps=0.12$ in table~\ref{tab:T1} and the comparison of section~\ref{sec:euler} for $\eps>0.11$ use $N=48$. For the reflection-symmetric models of figures~\ref{fig:FD} and~\ref{fig:FW} we integrate in the invariant subspace $A_{-n}=A_n$ of section~\ref{sec:manifold}.

\textbf{The initial condition.} The integration starts on the unstable manifold. We compute the series \eqref{eq:series} to order $J=16$ from \eqref{eq:block}, with the eigenvector \eqref{eq:a1}, and evaluate it at the value $\xi_0$ where the linear term has amplitude $\delta_0$, that is, $\xi_0=\delta_0/\max_n\abs{a_{n,1}}$. The integration starts at $\tau_0=\lambda_1^{-1}\ln\xi_0$ from $A_n(\tau_0)=\sum_{p=0}^{J}a_{n,p}\,\xi_0^p$. We take $\delta_0=10^{-3}$ for the tables, the map of figure~\ref{fig:FA} and the comparison of section~\ref{sec:euler}, and $\delta_0=10^{-6}$ for the time series of figures~\ref{fig:FD}, \ref{fig:FW}, \ref{fig:FWD} and~\ref{fig:FC}(\textit{a}), which start earlier. This point lies well inside the disc of convergence. At $\eps=0.1$ and $\kone=1.3$, for instance, $\xi_0=5.2\times10^{-4}$, the last term of the series is of order $10^{-47}$, and the residual of the invariance equation at $\xi_0$ is $3\times10^{-13}$, so the initial state lies on the unstable manifold to round-off. Changing $\delta_0$ from $10^{-6}$ to $10^{-2}$, $J$ from $16$ to $24$, or the tolerances below changes $\Peak$ there by less than $3\times10^{-14}$. Starting instead from the linear term alone ($J=1$), as is usual, shifts $\Peak$ by $5\times10^{-6}$ at $\delta_0=10^{-3}$, $4\times10^{-4}$ at $10^{-2}$ and $4\times10^{-2}$ at $10^{-1}$, an error that grows like $\delta_0^2$ (figure~\ref{fig:FS}(\textit{d})).

\textbf{The integrator.} We use the explicit Runge--Kutta pair of orders 9 and 8 of Verner, MATLAB's \texttt{ode89}, with relative and absolute tolerances $10^{-12}$ and $10^{-14}$. The integration is extended in steps of $2/\lambda_1$ until $\tau_{\mathrm{ret}}$ has been found and the solution has gone $4/\lambda_1$ past $\tau_*$. If $\Pi$ has no local maximum after $\tau_{\min}$, the window is closed once the carrier has recovered $99\%$ of the wave action it lost. The wave action is conserved to about $10^{-15}$.

\textbf{The first peak.} Along the solution, $\max_X\abs{u}$ and $\Pi$ are sampled every $0.005/\lambda_1$ from the dense output of the integrator. The maximum over $X$ is located on an oversampled Fourier grid, with at least $64$ points and $8$ per kept mode, and refined by Newton's method on $\partial_X\abs{u}^2=0$. The times $\tau_{\min}$ and $\tau_{\mathrm{ret}}$ of \eqref{eq:firstevent} are read from the samples of $\Pi$. The instant $\tau_*$ of the peak is then refined by Brent's method (MATLAB's \texttt{fminbnd}, with tolerance $10^{-11}$), and $\Peak$ is evaluated from an integration step that ends exactly at $\tau_*$, since the interpolant of the dense output is less accurate than the steps themselves. For the models of table~\ref{tab:T1} other than the Dysthe equation with the exact dispersion, our code closes the window of \eqref{eq:firstevent} instead where $\max_X\abs{u}-1$ falls below $0.6$ of its running maximum; at every entry of the table both rules give the same $\Peak$.

\textbf{The free surface.} The surface \eqref{eq:recon} is reconstructed at $\tau_*$ over one modulation period centred on the peak, on a grid with at least $64$ points per carrier wavelength, for $256$ carrier phases equally spaced in $[0,2\pi)$. The waves are cut at the zero up-crossings that a wave gauge would record; since the waves travel towards $+X$, these are the down-crossings of $\eta$ in $X$. Each crest and trough is refined by a parabola through the three grid values around it, and the maxima over the carrier phase by a golden-section search.

\section{The first orders of the recursion}\label{sec:firstorders}

This appendix works out the second and third orders of the recursion of section~\ref{sec:construction}. They make the construction concrete and help to check an implementation.

\subsection{Second order}\label{sec:closedforms}

The coefficients available below order two are $a_{0,0}=1$ and $a_{\pm1,1}$. A product of total order two therefore has one factor $A_0$ and two factors from the pair $n=\pm1$: either the conjugated factor is $\bar A_0$ and $l,m=\pm1$, or it is $\bar A_{\pm1}$ and one of $l$, $m$ is zero. In both cases $n=l+m-j$ is $-2$, $0$ or $2$.

The quintic sum, present only in the Stokes-corrected equation, is treated in the same way. A product of total order two again has two factors from the pair $n=\pm1$, and its other three factors are $A_0$ or $\bar A_0$. Either both factors from the pair are unconjugated, so that the conjugated factors are $\bar A_0^2$ and two of $l$, $m$, $m'$ are $\pm1$; or one of them is conjugated, $\bar A_{\pm1}$, and one of $l$, $m$, $m'$ is $\pm1$; or both are conjugated and $l=m=m'=0$. In every case $n=l+m+m'-j-j'$ is again $-2$, $0$ or $2$. The quintic sum carries no kernel, so each product enters with the number of ways its factors can be placed in the slots $(j,j')$ and $(l,m,m')$: two unconjugated factors in three ways if they are equal and six if they differ, one conjugated and one unconjugated factor in $2\times3=6$ ways, and two conjugated factors in one way if they are equal and two if they differ.

Reading the coefficients off \eqref{eq:galerkinxi},
\begin{equation}\label{eq:R2}
\begin{aligned}
  R_{2,2}&=T_{2,0}^{1,1}\,a_{1,1}^2+2T_{2,-1}^{1,0}\,\bar a_{-1,1}a_{1,1}
    +\gamma\bigl(3a_{1,1}^2+6\bar a_{-1,1}a_{1,1}+\bar a_{-1,1}^2\bigr),\\
  R_{0,2}&=2T_{0,0}^{1,-1}\,a_{1,1}a_{-1,1}+2T_{0,1}^{1,0}\abs{a_{1,1}}^2+2T_{0,-1}^{-1,0}\abs{a_{-1,1}}^2\\
    &\quad+\gamma\bigl(6a_{1,1}a_{-1,1}+6\abs{a_{1,1}}^2+6\abs{a_{-1,1}}^2+2\bar a_{1,1}\bar a_{-1,1}\bigr),\\
  R_{-2,2}&=T_{-2,0}^{-1,-1}\,a_{-1,1}^2+2T_{-2,1}^{-1,0}\,\bar a_{1,1}a_{-1,1}
    +\gamma\bigl(3a_{-1,1}^2+6\bar a_{1,1}a_{-1,1}+\bar a_{1,1}^2\bigr).
\end{aligned}
\end{equation}
Every other $R_{n,2}$ vanishes, and with it $a_{n,2}$.

For the NLS, where every $T$ is one, the blocks \eqref{eq:block} at $n=0$ and $n=\pm2$ give
\begin{equation}\label{eq:a2}
  a_{0,2}=\frac{4\ii}{\sin\thone}\,\ee{\ii\thone},\qquad
  a_{\pm2,2}=\frac{2}{\cos\thone}\,\ee{\ii\thone}.
\end{equation}

\subsection{Third order}

Below order three the available coefficients are $a_{0,0}$, $a_{\pm1,1}$, $a_{0,2}$ and $a_{\pm2,2}$, and the
products of total order three have $n=l+m-j$ odd with $\abs{n}\le3$: they force the harmonics $\pm1$ and
$\pm3$ and no others. The harmonic that is new at this order is built from the leading coefficients alone,
while $R_{1,3}$ also contains $a_{0,2}$,
\begin{equation}\label{eq:R33}
\begin{aligned}
  R_{3,3}&=2T_{3,0}^{2,1}\,a_{1,1}a_{2,2}+T_{3,-1}^{1,1}\,\bar a_{-1,1}a_{1,1}^2
  +2T_{3,-1}^{2,0}\,\bar a_{-1,1}a_{2,2}+2T_{3,-2}^{1,0}\,\bar a_{-2,2}a_{1,1}\\
  &\quad+\gamma\bigl(a_{1,1}^3+6\bar a_{-1,1}a_{1,1}^2+3\bar a_{-1,1}^2a_{1,1}+6a_{1,1}a_{2,2}\\
  &\qquad+6\bar a_{-1,1}a_{2,2}+6\bar a_{-2,2}a_{1,1}+2\bar a_{-1,1}\bar a_{-2,2}\bigr),\\
  R_{1,3}&=2T_{1,0}^{1,0}\,(a_{0,2}+\bar a_{0,2})\,a_{1,1}+2T_{1,0}^{-1,2}\,a_{-1,1}a_{2,2}
  +T_{1,1}^{1,1}\abs{a_{1,1}}^2a_{1,1}+2T_{1,-1}^{1,-1}\abs{a_{-1,1}}^2a_{1,1}\\
  &\quad+2T_{1,1}^{2,0}\,\bar a_{1,1}a_{2,2}+2T_{1,-1}^{0,0}\,\bar a_{-1,1}a_{0,2}
  +2T_{1,-2}^{-1,0}\,\bar a_{-2,2}a_{-1,1}\\
  &\quad+\gamma\bigl(6\abs{a_{1,1}}^2a_{1,1}+12\abs{a_{-1,1}}^2a_{1,1}+3a_{1,1}^2a_{-1,1}
  +6\abs{a_{1,1}}^2\bar a_{-1,1}\\
  &\qquad+3\abs{a_{-1,1}}^2\bar a_{-1,1}+6(a_{0,2}+\bar a_{0,2})\,a_{1,1}+6\bar a_{-1,1}a_{0,2}
  +2\bar a_{-1,1}\bar a_{0,2}\\
  &\qquad+6a_{-1,1}a_{2,2}+6\bar a_{1,1}a_{2,2}+6\bar a_{-2,2}a_{-1,1}+2\bar a_{1,1}\bar a_{-2,2}\bigr),
\end{aligned}
\end{equation}
and $R_{-3,3}$ and $R_{-1,3}$ are their mirror images. For the NLS
\begin{equation}\label{eq:a3}
  a_{\pm1,3}=-\frac{3-\ee{2\ii\thone}}{\sin^2\thone\cos\thone},\qquad
  a_{\pm3,3}=\frac{2}{\cos^2\thone}\,\ee{\ii\thone}.
\end{equation}

\section{The Akhmediev breather in Fourier space}\label{sec:ABfourier}

This appendix writes the Akhmediev breather, as given by \citet{chin_etal2015}, in our variables and computes its Fourier modes.

In the notation of \citet{chin_etal2015} the Akhmediev breather depends on a free parameter $a\in(0,\tfrac12)$, through which its modulation wavenumber is $\Omega=2(1-2a)^{1/2}$ and its growth rate is $\lambda=[8a(1-2a)]^{1/2}$ (their eqs.~5 and~6; their $\Omega$ is not our dispersion multiplier). With $\Omega=k_1$ and $\cos\thone=\tfrac12k_1$ this means $(1-2a)^{1/2}=\cos\thone$, $(2a)^{1/2}=\sin\thone$, $1-4a=\cos2\thone$ and $\lambda=\sin2\thone=\lambda_1$, and the breather reads
\begin{equation}\label{eq:abchin}
  \psi_{AB}(X,\tau)=\ee{\ii\tau}\,
  \frac{\cos2\thone\cosh\lambda_1\tau+\sin\thone\cos k_1X+\ii\sin2\thone\sinh\lambda_1\tau}
       {\sin\thone\cos k_1X-\cosh\lambda_1\tau},
\end{equation}
with its peak, $\abs{\psi_{AB}}=1+2\sin\thone$, at $X=\tau=0$. As $\tau\to-\infty$ it tends to the Stokes wave $-\ee{-2\ii\thone}\ee{\ii\tau}$ rather than to $\ee{\ii\tau}$. By the gauge invariance, $-\ee{2\ii\thone}\psi_{AB}$ is also a solution; it tends to $\ee{\ii\tau}$ and therefore lies on $\Wu$. In the rotating frame of \eqref{eq:galerkin},
\begin{equation}\label{eq:abtau}
  \sum_nA_n\ee{\ii nk_1X}=-\ee{2\ii\thone}\ee{-\ii\tau}\psi_{AB}
  =\frac{1+2\sin\thone\,\ee{2\ii\thone}\ee{\lambda_1\tau}\cos k_1X+\ee{4\ii\thone}\ee{2\lambda_1\tau}}
        {1-2\sin\thone\,\ee{\lambda_1\tau}\cos k_1X+\ee{2\lambda_1\tau}}.
\end{equation}
To first order in $\ee{\lambda_1\tau}$ this gives $A_{\pm1}=2\sin\thone\cos\thone\,\ee{\ii\thone}\ee{\lambda_1\tau}$, which is the first-order coefficient $a_{\pm1,1}=2\ee{\ii\thone}$ of \eqref{eq:a1} times $\varrho\,\ee{\lambda_1\tau}$, with $\varrho=\sin\thone\cos\thone$. The variable of the series \eqref{eq:series} is therefore
\begin{equation}\label{eq:xiab}
  \xi=\sin\thone\cos\thone\,\ee{\lambda_1\tau}=\varrho\,\ee{\lambda_1\tau}.
\end{equation}

Equation \eqref{eq:xiab} is not a new definition of $\xi$ but a shift of the origin of time. In section~\ref{sec:construction} the origin of $\tau$ is fixed by the normalisation $r=1$ of the eigenvector \eqref{eq:a1}, which gives $\xi=\ee{\lambda_1\tau}$, whereas the time in \eqref{eq:abchin} has its origin at the peak. The two times differ by $\lambda_1^{-1}\ln\varrho$, and \eqref{eq:xiab} writes our $\xi$ in the time of \eqref{eq:abchin}, the time that we use for the Akhmediev breather in the main text.

In terms of $\xi$, \eqref{eq:abtau} becomes
\begin{equation}\label{eq:abxi}
  \sum_nA_n\ee{\ii nk_1X}=
  \frac{1+2\xi\sec\thone\,\ee{2\ii\thone}\cos k_1X+\ee{4\ii\thone}\xi^2/\varrho^2}
       {1-2\xi\sec\thone\cos k_1X+\xi^2/\varrho^2}.
\end{equation}

The Fourier modes of the breather are obtained from the identity
\[
  \frac{1}{1-\nu\cos y}=\frac{1}{(1-\nu^2)^{1/2}}\sum_nq^{\abs{n}}\ee{\ii ny},\qquad q=\frac{1-(1-\nu^2)^{1/2}}{\nu},
\]
valid for $\abs{\nu}<1$. The denominator of \eqref{eq:abtau} is $2\ee{\lambda_1\tau}\cosh\lambda_1\tau\,(1-\nu\cos k_1X)$ with $\nu=\sin\thone\operatorname{sech}\lambda_1\tau$, and the Fourier modes are
\begin{equation}\label{eq:abmodes1}
  A_n=F\,q^{\abs{n}}\quad(n\ne0),\qquad A_0=F-\ee{2\ii\thone},\qquad
  F=\ee{2\ii\thone}\,\frac{2\cos^2\thone+\ii\sin2\thone\tanh\lambda_1\tau}{(1-\nu^2)^{1/2}},
\end{equation}
which are the amplitudes of \citet[eqs.~31 and~32]{chin_etal2015} multiplied by $-\ee{2\ii\thone}$.

In terms of $\xi$ the modes become algebraic. By \eqref{eq:xiab}, $\nu=2\varrho\sin\thone\,\xi/(\varrho^2+\xi^2)$, and
\begin{gather*}
  (1-\nu^2)^{1/2}=\frac{(\varrho^2+\ee{2\ii\thone}\xi^2)^{1/2}(\varrho^2+\ee{-2\ii\thone}\xi^2)^{1/2}}{\varrho^2+\xi^2},\\
  2\cos^2\thone+\ii\sin2\thone\tanh\lambda_1\tau=2\cos\thone\,\ee{-\ii\thone}\,\frac{\varrho^2+\ee{2\ii\thone}\xi^2}{\varrho^2+\xi^2},
\end{gather*}
so that $F$ and $q$ take the form \eqref{eq:abmodesxi}, with the principal square roots, which equal $\varrho$ at $\xi=0$. Every mode is therefore built from the two square roots $(\varrho^2+\ee{\pm2\ii\thone}\xi^2)^{1/2}$. For real $\xi$ they are complex conjugates of each other, so $q$ is real and $\abs{F}=2\cos\thone$ at all times: the modes $n\ne0$ share the phase of $F$ and their moduli decrease geometrically with ratio $q$.

\bibliographystyle{jfm}
\bibliography{references}

\end{document}